\documentclass[11pt,a4paper]{article}
\pdfoutput=1

\usepackage[utf8]{inputenc}
\usepackage{a4wide}
\usepackage{amsmath}
\usepackage{cite}
\usepackage{xcolor}
\usepackage{amssymb}
\usepackage{amsfonts}
\usepackage{booktabs}
\usepackage{makecell}
\usepackage{multirow}
\usepackage{pdflscape}
\usepackage{textcomp}
\usepackage{gensymb}
\usepackage{bigstrut}
\usepackage{array}
\usepackage{enumerate}
\usepackage[small,bf]{caption}
\usepackage{graphicx}
\usepackage{hyperref}
\usepackage{mathbbol}
\usepackage{amssymb}
\usepackage{appendix}
\usepackage{verbatim}
\usepackage{geometry}
\usepackage{makecell}
\usepackage{listings}
\usepackage{mathtools}
\usepackage{dsfont}
\usepackage{accents}
\usepackage{longtable}
\usepackage{colortbl}

\DeclareMathOperator{\diag}{diag}
\DeclareMathOperator{\tr}{Tr}
\DeclareMathOperator{\im}{Im}
\DeclareMathOperator{\re}{Re}
\def\gtap{\ \raisebox{-.4ex}{\rlap{\(\sim\)}} \raisebox{.4ex}{\(>\)}\ }
\def\ltap{\ \raisebox{-.4ex}{\rlap{\(\sim\)}} \raisebox{.4ex}{\(<\)}\ }

\newcommand{\id}{\mathds{1}}

\newcommand*{\belowrulesepcolor}[1]{%
  \noalign{%
    \kern-\belowrulesep
    \begingroup
      \color{#1}%
      \hrule height\belowrulesep
    \endgroup
  }%
}
\newcommand*{\aboverulesepcolor}[1]{%
  \noalign{%
    \begingroup
      \color{#1}%
      \hrule height\aboverulesep
    \endgroup
    \kern-\aboverulesep
  }%
}
\definecolor{light-gray}{gray}{0.9}
\definecolor{lighter-gray}{gray}{0.95}

\newcommand\tildepar[1]{\accentset{\displaystyle\resizebox{2pt}{3pt}{(}\text{\texttildelow}\resizebox{2pt}{3pt}{)}}{#1}}

\allowdisplaybreaks
\numberwithin{equation}{section}

\begin{document}

%%%%%%%%%%%%%%%%%%%%%%%
\begin{titlepage}

\vspace*{-15mm}
\begin{flushright}
SISSA 09/2021/FISI \\
IPMU21-0009 \\
CFTP/21-003
\end{flushright}
\vspace*{5mm}

\begin{center}
{\bf\LARGE Fermion Mass Hierarchies,
Large Lepton Mixing\\[3mm]
and Residual Modular Symmetries
}\\[8mm]

P.~P.~Novichkov\(^{\,a,}\)\footnote{E-mail: \texttt{pavel.novichkov@sissa.it}},
J.~T.~Penedo\(^{\,b,}\)\footnote{E-mail: \texttt{joao.t.n.penedo@tecnico.ulisboa.pt}},
S.~T.~Petcov\(^{\,a,c,}\)\footnote{Also at
Institute of Nuclear Research and Nuclear Energy,
Bulgarian Academy of Sciences, 1784 Sofia, Bulgaria.}\\
 \vspace{5mm}
\(^{a}\)\,{\it SISSA/INFN, Via Bonomea 265, 34136 Trieste, Italy} \\
\vspace{2mm}
\(^{b}\)\,{\it CFTP, Departamento de Física, Instituto Superior Técnico, Universidade de Lisboa,\\
Avenida Rovisco Pais 1, 1049-001 Lisboa, Portugal} \\
\vspace{2mm}
\(^{c}\)\,{\it Kavli IPMU (WPI), University of Tokyo, 5-1-5 Kashiwanoha, 277-8583 Kashiwa, Japan}
\end{center}
\vspace{2mm}

\begin{abstract}
In modular-invariant models of flavour,
hierarchical fermion mass matrices may arise
solely due to the proximity of the modulus \(\tau\)
to a point of residual symmetry.
This mechanism does not require flavon fields, and modular weights are not analogous to Froggatt-Nielsen charges.
Instead, we show that hierarchies depend on the decomposition of field representations under the residual symmetry group.
We systematically go through the 
possible fermion field representation choices which may yield hierarchical structures in the vicinity of symmetric points,
for the four smallest finite modular groups, isomorphic to \(S_3\), \(A_4\), \(S_4\), and \(A_5\), as well as for their double covers. 
We find a restricted set of pairs of representations for which the 
discussed mechanism may produce viable fermion (charged-lepton and quark) mass hierarchies.
We present two lepton flavour models in which the charged-lepton 
mass hierarchies are naturally obtained, while lepton mixing is somewhat fine-tuned.
After formulating the conditions for obtaining a viable lepton mixing matrix in the symmetric limit, we construct a model in which both the charged-lepton and neutrino sectors are free from fine-tuning.
\end{abstract}

\end{titlepage}
\setcounter{footnote}{0}
%%%%%%%%%%%%%%%%%%%%%%%
%

\tableofcontents
\vskip 1cm

\vfill
\clearpage

%%%%%%%%%%%%%%%%%%%%%%%
\section{Introduction}
\label{sec:intro}
%%%%%%%%%%%%%%%%%%%%%%%
%
Understanding the origins of
flavour in both the quark and lepton sectors, i.e., 
the origins of the patterns of quark masses and mixing, 
of the charged-lepton and neutrino masses, of neutrino mixing 
and of the CP violation in the two sectors is one of the most challenging 
unresolved fundamental problems in particle physics~\cite{Feruglio:2015jfa}.%
\footnote{
 ``Asked what single mystery, if he could choose, he would like to see 
 solved in his lifetime, Weinberg doesn't have to think for long: 
 he wants to be able to explain the observed pattern of quark 
 and lepton masses.''
 From {\it Model Physicist}, CERN Courier, 13 October 2017.
}

Within the reference three neutrino mixing scheme,
the lepton flavour problem consists of three 
basic elements or sub-problems, namely, understanding:
\begin{enumerate}[i)]
  \item the origin of the hierarchical pattern of charged-lepton 
masses: \(m_e \ll m_\mu \ll m_\tau\),  \(m_e/m_\mu \simeq 1/200\),  
\(m_\mu/ m_\tau \simeq 1/17\);
  \item why neutrino masses \(m_{\nu_j}\) are much smaller than the masses
of charged leptons and quarks, 
\(m_{\nu_j} \lll m_{\ell,q}\), \(q=u,c,t,d,s,b\) and \(\ell=e,\mu,\tau\), with  
\(m_{\nu_j}\ltap 0.5\) eV, \(m_{\ell}\geq 0.511\) MeV, \(m_{q}\gtap 2\) MeV; 
  \item the origins of the patterns of neutrino mixing 
of 2 large and 1 small angles, and of the two independent 
neutrino mass squared differences,
\(\Delta m^2_{21} \ll |\Delta m^2_{31}|\) with
\(\Delta m^2_{21}/|\Delta m^2_{31}| \simeq 1/30\), where 
\(\Delta m^2_{ij}\equiv m^2_i - m^2_j\).
\end{enumerate}

 Each of these three sub-problems  
is by itself a formidable problem.
As a consequence, individual solutions to each of 
them have been proposed.
The hierarchical pattern of charged-lepton 
masses can most naturally be understood 
within the Froggatt-Nielsen mechanism based 
on the \(U(1)_{\rm FN}\) flavour symmetry~\cite{Froggatt:1978nt} and its extensions. 
The enormous disparity between the neutrino masses
and the masses of the charged leptons and quarks 
can be understood within the seesaw 
or radiative models of neutrino mass generation 
or else employing the Weinberg 
effective operator idea~\cite{Weinberg:1979sa}  
(for a concise review see, e.g.,~\cite{Petcov:2019fmk}).
All these approaches lead naturally 
to massive Majorana neutrinos.
Arguably the most elegant and natural 
explanation of the observed pattern of 
neutrino (or lepton) mixing of two large and one small 
mixing angles is obtained within the non-Abelian 
discrete symmetry approach to the problem (see, e.g.,~\cite{Altarelli:2010gt,Ishimori:2010au,King:2014nza,Tanimoto:2015nfa,Petcov:2017ggy}). 

In the case of the quark sector, the flavour 
problem similarly has two basic sub-problems,
namely, understanding:
\begin{enumerate}[i)]
  \item  the origins of the hierarchies of the masses of the charge  
\(2/3\) and of the charge (\(-1/3\)) quarks;
  \item  the origins of the relatively small values of the three 
quark mixing angles.
\end{enumerate} 
The most natural qualitative solution 
of these two problems  
is arguably provided by the Froggatt-Nielsen 
approach~\cite{Froggatt:1978nt}, 
although the approach based on non-Abelian 
discrete symmetries has been applied to the 
quark flavour problem as well.
Solutions to the two flavour problems 
within the theories with extra dimensions 
have also been proposed.% 
\footnote{A rather comprehensive 
discussion of the past proposed 
approaches to the lepton and quark flavour problems 
can be found in the review 
article~\cite{Feruglio:2015jfa}.
}

 The specific solutions to the individual 
lepton flavour sub-problems listed above 
become problematic when applied to the 
sub-problems they were not intended to solve.
The seesaw and the radiative neutrino mass 
models do not lead to understanding 
of the origin of the pattern of 
neutrino mixing without additional input,
consisting typically of imposing 
specific additional symmetries
(of GUT or flavour type) on the relevant 
constructions. Within the Froggatt-Nielsen 
approach one most naturally obtains 
small values of the three neutrino mixing
angles and while the charged-lepton and quark 
mass hierarchies can be qualitatively understood 
within this approach, 
the specific predictions suffer 
from relatively large uncertainties. 
The symmetry breaking in the lepton and quark flavour models 
based on non-Abelian discrete symmetries
is impressively cumbersome: it requires 
the introduction of a plethora of 
`flavon' scalar fields having 
elaborate potentials, which in turn require 
the introduction of large shaping symmetries 
to ensure the requisite breaking of the symmetry 
leading to correct mass and mixing patterns.

 There have been also attempts to 
 make progress, e.g., on the 
lepton flavour problem by combining the 
proposed `solutions' of the three related 
sub-problems. In these combined approaches it is difficult, 
if not impossible, to avoid the drawbacks of each 
of the `ingredient' sub-problem `solutions'. 
In some cases this can be achieved at the 
cost of severe fine-tuning.
Thus, a universal, elegant, natural 
and viable theory of flavour 
that is free from undesired drawback features 
is still lacking. Constructing such a theory would be a major breakthrough in particle physics.

The unsatisfactory status of the flavour
problem and the remarkable progress 
made in the studies of neutrino 
oscillations (see, e.g., \cite{Tanabashi:2018oca}),
which began 22 years ago with the discovery 
of oscillations of the atmospheric \(\nu_{\mu}\) and 
\(\bar{\nu}_{\mu}\) by the Super-Kamiokande experiment 
\cite{Fukuda:1998mi} and lead, in particular, 
to the determination of the pattern of 
neutrino mixing, stimulated renewed attempts to seek 
alternative viable approaches to the lepton 
as well as to the quark flavour problems.
A step in this direction was made in \cite{Feruglio:2017spp}
where the idea of using modular invariance as a flavour 
symmetry was put forward. 
This new original approach based on modular invariance
opened up a new promising direction in the studies of the flavour 
problem and correspondingly in flavour model building.

 The main feature of the approach proposed in \cite{Feruglio:2017spp}  
is that the elements of the Yukawa coupling 
and fermion mass matrices in the Lagrangian of the theory are
modular forms of a certain level \(N\) which 
are functions of a single complex scalar field \(\tau\)
-- the modulus -- and have specific transformation properties 
under the action of the modular group.  
In addition, both the couplings and the matter fields
(supermultiplets) are assumed to
transform in representations of an inhomogeneous (homogeneous) 
finite modular group \(\Gamma^{(\prime)}_N\). 
For \(N\leq 5\), the finite modular groups \(\Gamma_N\) 
are isomorphic to the permutation groups 
\( S_3\), \( A_4\), \( S_4\) and \( A_5\)
 (see, e.g., \cite{deAdelhartToorop:2011re}), 
while the groups \(\Gamma^\prime_N\) are isomorphic to the double
covers of the indicated permutation groups,
\(S^\prime_3 \equiv S_3\), \(A^\prime_4 \equiv T^\prime\), 
\(S^\prime_4\) and \(A^\prime_5\). These discrete groups 
are widely used in flavour model building.
The theory is assumed to possess the modular symmetry described by the finite modular group \(\Gamma^{(\prime)}_N\), which plays the role of a flavour symmetry.
In the simplest class of such models, 
the VEV of the modulus \(\tau\) is the only source of flavour 
symmetry breaking, such that no flavons are needed. 
Another appealing feature of the proposed framework is that 
the VEV of \(\tau\) can also be the only source of breaking of the 
CP symmetry \cite{Novichkov:2019sqv}.
When the  flavour symmetry is broken, 
the elements of the Yukawa coupling and fermion mass matrices
get fixed, and a certain flavour structure arises. 
As a consequence of the modular symmetry, 
in the lepton sector,  for example, 
the charged-lepton and neutrino masses, 
neutrino mixing and the leptonic CPV phases
are simultaneously determined  
in terms of a limited number of coupling constant parameters. 
This together with the fact that they are also functions of 
a single complex VEV -- that of the modulus \(\tau\) --
leads to experimentally testable 
correlations between, e.g., the neutrino mass and 
mixing observables. Models of flavour based on modular invariance 
have then an increased predictive power.

The modular symmetry approach to the flavour problem 
has been widely implemented so far primarily 
in theories with global (rigid) supersymmetry.  
Within the SUSY framework, modular invariance is 
assumed to be a feature of the Kähler potential 
and the superpotential of the theory.%
\footnote{Possible non-minimal additions to the Kähler potential, 
compatible with the modular symmetry, may jeopardise the predictive power 
of the approach~\cite{Chen:2019ewa}.
This problem is the subject of ongoing research.
}
Bottom-up modular invariance approaches to the lepton 
flavour problem have been 
exploited first using the groups
 \(\Gamma_3 \simeq A_4\)~\cite{Feruglio:2017spp,Criado:2018thu}, 
\(\Gamma_2 \simeq S_3\)~\cite{Kobayashi:2018vbk}, 
\(\Gamma_4 \simeq S_4\)~\cite{Penedo:2018nmg}.
After the first studies, the interest in the approach grew significantly 
and models based on the groups 
\(\Gamma_4 \simeq S_4\)~\cite{Novichkov:2018ovf, Kobayashi:2019mna, Okada:2019lzv, Kobayashi:2019xvz, Gui-JunDing:2019wap, Wang:2019ovr,  Wang:2020dbp, Gehrlein:2020jnr},
\(\Gamma_5 \simeq A_5\)~\cite{Novichkov:2018nkm, Ding:2019xna, Gehrlein:2020jnr},
\(\Gamma_3 \simeq A_4\)~\cite{Kobayashi:2018scp, Novichkov:2018yse, Nomura:2019jxj, Nomura:2019yft, Ding:2019zxk, Okada:2019mjf, Nomura:2019lnr, Asaka:2019vev, Gui-JunDing:2019wap, Zhang:2019ngf, Nomura:2019xsb, Kobayashi:2019gtp, Wang:2019xbo, Abbas:2020vuy, Okada:2020dmb, Ding:2020yen, Behera:2020sfe, Nomura:2020opk, Nomura:2020cog, Behera:2020lpd, Asaka:2020tmo, Nagao:2020snm, Hutauruk:2020xtk},
\(\Gamma_2 \simeq S_3\)~\cite{Okada:2019xqk, Mishra:2020gxg}
and \(\Gamma_7 \simeq PSL(2,\mathbb{Z}_7)\)~\cite{Ding:2020msi}
have been constructed and extensively studied.
Similarly, attempts have been made to construct viable models of 
quark flavour~\cite{Okada:2018yrn} and of quark-lepton unification~\cite{Kobayashi:2018wkl,Okada:2019uoy,Kobayashi:2019rzp,Lu:2019vgm,Abbas:2020qzc,Okada:2020rjb,Du:2020ylx,Zhao:2021jxg,Chen:2021zty}.
The formalism of the interplay of modular and gCP symmetries
has been developed and first applications made in~\cite{Novichkov:2019sqv}. 
It was explored further in~\cite{Kobayashi:2019uyt,Okada:2020brs,Yao:2020qyy,Wang:2021mkw},
as was the possibility of coexistence 
of multiple moduli~\cite{deMedeirosVarzielas:2019cyj, King:2019vhv, deMedeirosVarzielas:2020kji, Ding:2020zxw}, 
considered first phenomenologically 
in~\cite{Novichkov:2018ovf, Novichkov:2018yse}.
Such bottom-up analyses are expected to eventually connect with top-down 
results~\cite{Kobayashi:2018rad, Kobayashi:2018bff, deAnda:2018ecu, Baur:2019kwi, Kariyazono:2019ehj, Baur:2019iai, Nilles:2020nnc, Kobayashi:2020hoc, Abe:2020vmv, Ohki:2020bpo, Kobayashi:2020uaj, Nilles:2020kgo,Kikuchi:2020frp, Nilles:2020tdp, Kikuchi:2020nxn, Baur:2020jwc, Ishiguro:2020nuf, Nilles:2020gvu, Ishiguro:2020tmo, Hoshiya:2020hki, Baur:2020yjl, Kikuchi:2021ogn}
based on ultraviolet-complete theories.
While the aforementioned finite quotients~\(\Gamma_N\) of the modular
group have been widely used in the literature to construct modular-invariant
models of flavour from the bottom-up perspective, top-down
constructions typically lead to their double covers~\(\Gamma'_N\) (see,
e.g.,~\cite{Ferrara:1989qb,Baur:2019kwi,Baur:2019iai,Nilles:2020nnc}).
The formalism of such double covers has been 
developed and viable flavour models constructed in
Refs.~\cite{Liu:2019khw},~\cite{Novichkov:2020eep,Liu:2020akv}
and~\cite{Wang:2020lxk,Yao:2020zml} for the cases  
of \(\Gamma'_3 \simeq T'\), \(\Gamma'_4 \simeq S'_4\) and 
\(\Gamma'_5 \simeq A'_5\), respectively.

In almost all phenomenologically 
viable flavour models based on modular 
 invariance constructed so far
the hierarchy of the charged-lepton and quark 
masses is obtained  by fine-tuning some of the 
constant parameters present in the models.%
\footnote{By fine-tuning we refer to either i) high sensitivity of observables to model parameters or ii) unjustified hierarchies between parameters which are introduced in the model on an equal footing.}
Perhaps, the only notable exceptions are Refs.~\cite{Criado:2019tzk,King:2020qaj}, in which modular weights are used as Froggatt-Nielsen charges, and additional scalar fields of non-zero modular weights play the role of flavons.

In the present article we develop the formalism 
that allows to construct models in which 
the fermion (e.g.~charged-lepton and quark) mass hierarchies 
follow solely from the properties of the
modular forms present in the fermion mass matrices, 
thus avoiding the fine-tuning without the need to introduce extra fields.
We consider theories described by a modular group \(\Gamma'_N\) with \(N\leq 5\) (which encompasses the unprimed \(\Gamma_N\)).
It was noticed in \cite{Novichkov:2018ovf} 
and further exploited in 
\cite{Novichkov:2018yse,Novichkov:2018nkm,Okada:2020brs} that  
for the three fixed points of the VEV of \(\tau\) 
in the modular group fundamental domain,
\(\tau_\text{sym} = i\), \(\tau_\text{sym} = 
\omega \equiv \exp(i\,2\pi/ 3)
= -\,1/2 + i\sqrt{3}/2\) (the `left cusp'),
and \(\tau_\text{sym} = i\infty\),
the theories based on the \(\Gamma_N\) 
invariance have respectively 
\(\mathbb{Z}^{S}_2\),
\(\mathbb{Z}^{ST}_3\), and 
\(\mathbb{Z}^{T}_N\) residual symmetries. 
In the case of the double cover groups \(\Gamma^\prime_N\),
the \(\mathbb{Z}^{S}_2\) residual symmetry is replaced by 
the \(\mathbb{Z}^{S}_4\) 
and there is an additional \(\mathbb{Z}_2^R\) symmetry
that is unbroken for any value of \(\tau\)
(see~\cite{Novichkov:2020eep} for further details).
The indicated residual symmetries play a crucial role 
in our analysis.

The fermion mass matrices are strongly constrained
in the points of residual symmetries. 
This suggests that fine-tuning could be avoided in the vicinity of these points
if the charged-lepton and quark 
mass hierarchies follow from the properties 
of the modular forms present in the corresponding 
fermion mass matrices%
\footnote{
    In the case of \(\tau_{\text{sym}} = i \infty\) this idea is related to the fact that Yukawa couplings may become suppressed in the limit of large \(\im \tau\), which was originally noticed in the context of string theory.
    More specifically, the Yukawa couplings of twisted fields in heterotic orbifold models were shown to be exponentially suppressed by \(d^2 \propto \im \tau\) (cf.~eq.~\eqref{eq:q}), where \(d\) is the distance between the fixed points to which the fields are attached \cite{Hamidi:1986vh,Dixon:1986qv} (see also \cite{Lauer:1989ax,Lauer:1990tm}).
This mechanism was suggested as a possible origin of the observed hierarchies of quark and lepton masses (see, e.g., \cite{Ibanez:1986ka,Casas:1992zt}).
  }
rather than being determined by the values of
the accompanying constants also present    
in the matrices. 
Relatively small deviations of the modulus VEV 
from the symmetric point might also be needed 
to ensure the breaking of the CP symmetry~\cite{Novichkov:2019sqv}.

We note that in~\cite{Okada:2020ukr}
flavour models in the vicinity of the residual symmetry 
fixed points, \(\tau_\text{sym}=i, \omega, i\infty\), have been investigated
within the modular invariant \(A_4\) framework (\(N=3\)). 
The authors find viable lepton (quark) 
flavour models in the vicinity of each of three 
residual symmetry values of \(\tau_\text{sym}\) (of \(\tau_\text{sym} = i\)), 
in which the mixing arises seemingly without fine-tuning.
At the same time, the charged-lepton and quark 
mass hierarchies are obtained by fine-tuning the values 
of the constants present in the respective mass matrices. 

\vskip 2mm

The aim of this study is to investigate
the possibility of obtaining fermion mass hierarchies
-- and, in models of lepton flavour, large mixing -- without fine-tuning.
The article is structured as follows.
After introducing the necessary tools in section~\ref{sec:framework}, we describe how one can naturally generate hierarchical mass patterns in the vicinity of symmetric points in section~\ref{sec:theory}. In section~\ref{sec:decomp}, the role of decompositions under the residual symmetry groups is highlighted. We perform a systematic scan of attainable hierarchical patterns for \(N\leq 5\), the results of which are reported in section~\ref{sec:hierarchies}. The analysis of two promising lepton flavour models in section~\ref{sec:models_inf} motivates the discussion, in section~\ref{sec:natural}, of necessary conditions to avoid fine-tuned leptonic mixing. We are then driven to a subset of viable models, the most promising of which is explored in section~\ref{sec:models_S4_omega}. We summarise our results and conclude in section~\ref{sec:conclusions}.

\vfill
\clearpage

%%%%%%%%%%%%%%%%%%%%%%%
\section{Framework}
\label{sec:framework}
%%%%%%%%%%%%%%%%%%%%%%%
%
%%%%%%%%%%%%%%%%%%%%%%%
\subsection{Modular symmetries as flavour symmetries}
\label{sec:primer}
%%%%%%%%%%%%%%%%%%%%%%%
We start by briefly reviewing the modular invariance approach to flavour. 
In this supersymmetric (SUSY) framework, one introduces a chiral superfield, the modulus~\(\tau\), transforming non-trivially under the modular group \(\Gamma \equiv SL(2, \mathbb{Z})\). The group~\(\Gamma\) is generated by the matrices
\begin{equation}
  \label{eq:STR_def}
  S =
  \begin{pmatrix}
    0 & 1 \\ -1 & 0
  \end{pmatrix}
  \,, \quad
  T =
  \begin{pmatrix}
    1 & 1 \\ 0 & 1
  \end{pmatrix}
  \,, \quad
  R =
  \begin{pmatrix}
    -1 & 0 \\ 0 & -1
  \end{pmatrix}\,,
\end{equation}
obeying \(S^2 = R\), \((ST)^3 = R^2 = \id\), and \(RT = TR\).
Elements \(\gamma\) of the modular group act on \(\tau\) via fractional linear transformations,
\begin{equation}
  \label{eq:tau_mod_trans}
  \gamma =
  \begin{pmatrix}
    a & b \\ c & d
  \end{pmatrix}
  \in \Gamma : \quad
  \tau \to \gamma \tau = \frac{a\tau + b}{c\tau + d} \,,
\end{equation}
while matter superfields transform as `weighted' multiplets~\cite{Ferrara:1989bc,Ferrara:1989qb,Feruglio:2017spp},
\begin{equation}
  \label{eq:psi_mod_trans0}
  \psi_i \to (c\tau + d)^{-k} \, \rho_{ij}(\gamma) \, \psi_j \,,
\end{equation}
where \(k \in \mathbb{Z}\) is the so-called modular weight%
\footnote{While we restrict ourselves to integer \(k\), it is also possible for weights to be fractional \cite{Dixon:1989fj,Ibanez:1992hc,Olguin-Trejo:2017zav,Nilles:2020nnc}.}
and \(\rho\) is a unitary representation of~\(\Gamma\).

In using modular symmetry as a flavour symmetry, an integer level \(N \geq 2\) is fixed and one assumes that \(\rho(\gamma) = \id\) for elements \(\gamma\) of the principal congruence subgroup
\begin{equation}
  \label{eq:congr_subgr}
  \Gamma(N) \equiv
  \left\{
    \begin{pmatrix}
      a & b \\ c & d
    \end{pmatrix}
    \in SL(2, \mathbb{Z}), \,
    \begin{pmatrix}
      a & b \\ c & d
    \end{pmatrix}
    \equiv
    \begin{pmatrix}
      1 & 0 \\ 0 & 1
    \end{pmatrix}
    (\text{mod } N)
  \right\}\,.
\end{equation}
Hence, \(\rho\) is effectively a representation of the (homogeneous) finite modular group \(\Gamma_N' \equiv \Gamma \, \big/ \, \Gamma(N) \simeq SL(2, \mathbb{Z}_N)\). For \(N\leq 5\), this group admits the presentation
\begin{equation}
  \label{eq:hom_fin_mod_group_pres}
  \Gamma'_N = \left\langle S, \, T, \, R \mid S^2 = R, \, (ST)^3 = \id, \, R^2 = \id, \, RT = TR, \, T^N = \id \right\rangle\,.
\end{equation}

The (lowest component of the) modulus~\(\tau\) acquires a VEV which is restricted to the upper half-plane and plays the role of a spurion, parameterising the breaking of modular invariance. Additional flavon fields are not required, and we do not consider them here. Since~\(\tau\) does not transform under the \(R\) generator, a \(\mathbb{Z}_2^R\) symmetry is preserved in such scenarios.
If also matter fields transform trivially under \(R\), one may identify the matrices \(\gamma\) and \(-\gamma\), thereby restricting oneself to the inhomogeneous modular group~\(\overline{\Gamma} \equiv PSL(2, \mathbb{Z}) \equiv SL(2, \mathbb{Z}) \, / \, \mathbb{Z}_2^R\). 
In such a case, \(\rho\) is effectively a representation of a smaller (inhomogeneous) finite modular group \(\Gamma_N \equiv \Gamma \, \big/ \left\langle \, \Gamma(N) \cup \mathbb{Z}_2^R \, \right\rangle\). For \(N\leq 5\), this group admits the presentation
\begin{equation}
  \label{eq:inhom_fin_mod_group_pres}
  \Gamma_N = \left\langle S, \, T \mid S^2 = \id, \, (ST)^3 = \id, \, T^N = \id \right\rangle \,.
\end{equation}
In general, however, \(R\)-odd fields may be present in the theory and \(\Gamma\) and \(\Gamma_N'\) are then the relevant symmetry groups.
As shown in Table~\ref{tab:fin_mod_group}, the finite modular groups \(\Gamma_N\) and \(\Gamma'_N\) are isomorphic to permutation groups and to their double covers for small \(N\).
%
%%%%%%%%%%%%%%%%%%%%%%%
\begin{table}
  \centering
  \begin{tabular}{ccccc}
    \toprule
    \(N\) & 2 & 3 & 4 & 5 \\
    \midrule
    \(\Gamma_N\) & \(S_3\) & \(A_4\) & \(S_4\) & \(A_5\) \\
    \(\Gamma'_N\) & \(S_3\) & \(A'_4 \equiv T'\) & \(S'_4 \equiv SL(2, \mathbb{Z}_4)\) & \(A'_5 \equiv SL(2, \mathbb{Z}_5)\) \\
    \midrule
    \(\dim \mathcal{M}_k(\Gamma(N))\) & \(k/2 + 1\) & \(k+1\) & \(2k + 1\) & \(5k + 1\) \\
    \bottomrule
  \end{tabular}
  \caption{Finite modular groups and dimensionality of the corresponding spaces of modular forms, for \(N \leq 5\).
  Note that for \(N = 2\) only even-weighted modular forms exist.}
  \label{tab:fin_mod_group}
\end{table}
%%%%%%%%%%%%%%%%%%%%%%%
%
Group-theoretical results for the \(\Gamma_N\) groups are collected in appendix B of~\cite{Novichkov:2019sqv}, while for the double cover groups \(\Gamma'_N\) they can be found in Refs.~\cite{Liu:2019khw,Novichkov:2020eep,Wang:2020lxk}.

\vskip 2mm

Finally, to understand how modular symmetry may constrain the Yukawa couplings and mass structures of a model in a predictive way, we turn to the Lagrangian -- which for an \(\mathcal{N} = 1\) global supersymmetric theory is given by
\begin{equation}
  \mathcal{L} = \int \text{d}^2 \theta \, \text{d}^2 \bar{\theta} \, K(\tau,\bar{\tau}, \psi_I, \bar{\psi}_I)
  + \left[ \, \int \text{d}^2 \theta \, W(\tau,\psi_I) + \text{h.c.} \right] \,.
\end{equation}
Here \(K\) is the Kähler potential, while the superpotential \(W\) can be expanded in powers of matter superfields \(\psi_I\),
\begin{equation}
  \label{eq:W_series}
  W(\tau, \psi_I) = \sum \left( \vphantom{\sum} Y_{I_1 \ldots I_n}(\tau) \, \psi_{I_1} \ldots \psi_{I_n} \right)_{\mathbf{1}} \,,
\end{equation}
where one has summed over all possible field combinations and independent singlets of the finite modular group.
By requiring the invariance of the superpotential under modular transformations,%
\footnote{
In theories of supergravity \(W\) transforms under the modular symmetry with a certain weight~\cite{Ferrara:1989bc,Ferrara:1989qb}, shifting the required weights \(k\) of the modular forms.
}
one finds that the field couplings \(Y_{I_1 \ldots I_n}(\tau)\) have to be modular forms of level \(N\). These are severely constrained holomorphic functions of~\(\tau\), which under modular transformations obey
\begin{equation}
  \label{eq:Y_mod_trans}
  Y_{I_1 \ldots I_n}(\tau) \,\xrightarrow{\gamma}\, Y_{I_1 \ldots I_n}(\gamma \tau) = (c\tau + d)^{k} \rho_Y(\gamma) \,Y_{I_1 \ldots I_n}(\tau) \,.
\end{equation}
Modular forms carry weights \(k = k_{I_1} + \ldots + k_{I_n}\) and furnish unitary representations \(\rho_Y\) of the finite modular group such that \(\rho_Y \,\otimes\, \rho_{I_1} \,\otimes \ldots \otimes\, \rho_{I_n} \supset \mathbf{1}\).
Non-trivial modular forms of a given level exist only for \(k \in \mathbb{N}\), span finite-dimensional linear spaces~\(\mathcal{M}_{k}(\Gamma(N))\), and can be arranged into multiplets of \(\Gamma^{(\prime)}_N\). The fact that these spaces have low dimensionalities for small values of~\(k\) and \(N\) (as shown in Table~\ref{tab:fin_mod_group}) is at the root of the predictive power of the described setup, since only a restricted number of \(\tau\)-dependent Yukawa textures are allowed in the superpotential.

Note that modular forms are functions of \(\tau\) and are thus invariant under \(R\). In order to compensate the \((-1)^{k}\) factor in eq.~\eqref{eq:Y_mod_trans}, odd-weighted forms must furnish representations with~\(\rho_Y(R) = -\id\) (we use hats to denote such representations). For even-weighted modular forms, one has instead \(\rho_Y(R) = \id\).

%%%%%%%%%%%%%%%%%%%%%%%
\subsection{Residual symmetries}
\label{sec:residual}
%%%%%%%%%%%%%%%%%%%%%%%
The breakdown of modular symmetry is parameterised by the VEV of the modulus and there is no value of \(\tau\) which preserves the full symmetry. Nevertheless, at certain so-called symmetric points \(\tau = \tau_\text{sym}\) the modular group is only partially broken, with the unbroken generators giving rise to residual symmetries.
Recall that the \(R\) generator is unbroken for any value of \(\tau\), so that a \(\mathbb{Z}_2^R\) symmetry is always preserved.

%%%%%%%%%%%%%%%%
\begin{figure}[ht!]
  \centering
  \includegraphics[width=0.5\textwidth]{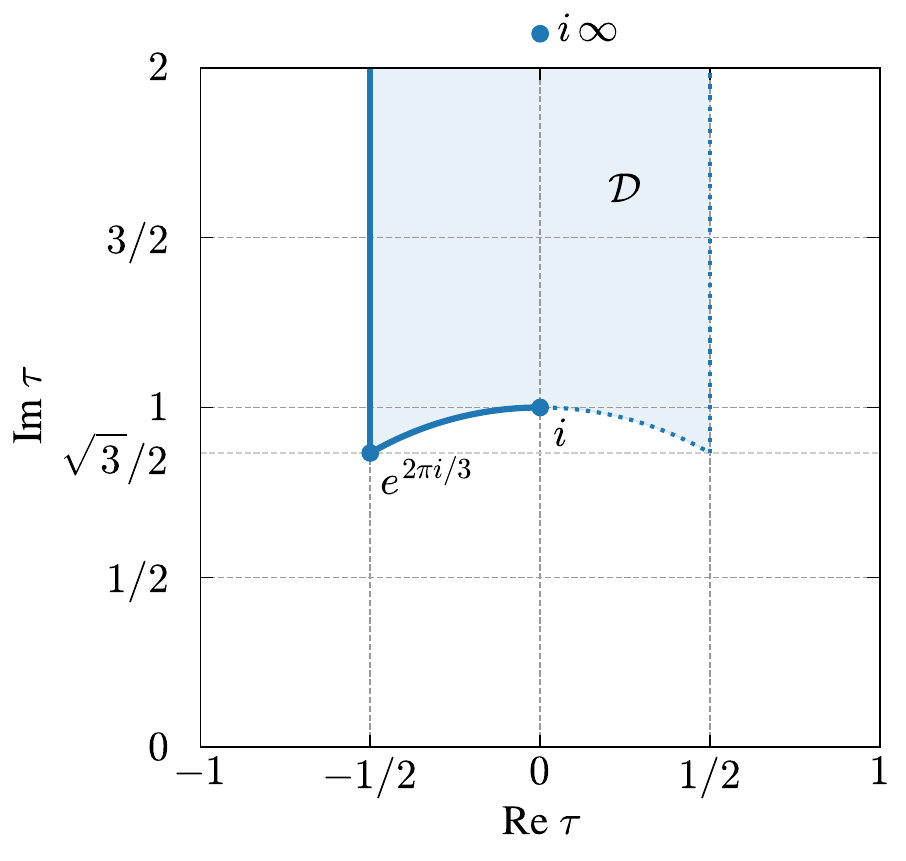}
  \caption{The fundamental domain \(\mathcal{D}\) of the modular group \(\Gamma\) and its three symmetric points \(\tau_\text{sym} = i\, \infty, i, \omega\). The value of \(\tau\) can always be restricted to \(\mathcal{D}\) by a suitable modular transformation. Figure from Ref.~\cite{Novichkov:2020eep}.}
  \label{fig:fund_domain}
\end{figure}
%%%%%%%%%%%%%%%%

The fundamental domain \(\mathcal{D}\) of the modular group is shown in Figure~\ref{fig:fund_domain}, along with its symmetric points. There are only three inequivalent symmetric points, namely~\cite{Novichkov:2018ovf}:
\begin{itemize}
\item \(\tau_\text{sym} = i \infty\), invariant under \(T\), preserving \(\mathbb{Z}_N^T \times \mathbb{Z}_2^R\);
\item \(\tau_\text{sym} = i\), invariant under \(S\), preserving \(\mathbb{Z}_4^S \) (recall that \(S^2 = R\)); and
\item \(\tau_\text{sym} = \omega \equiv \exp (2\pi i / 3)\), `the left cusp', invariant under \(ST\), preserving \(\mathbb{Z}_3^{ST} \times \mathbb{Z}_2^R\).
\end{itemize}

Finally, it is worth noting that these symmetric values preserve the CP symmetry of a CP- and modular-invariant theory (e.g.~a modular theory where the couplings satisfy a reality condition)~\cite{Novichkov:2019sqv,Novichkov:2020eep}. A \(\mathbb{Z}^\text{CP}_2\) symmetry is preserved for \(\re\tau =0\) or for \(\tau\) lying on the border of \(\mathcal{D}\), but is broken at generic values of \(\tau\).

%%%%%%%%%%%%%%%%%%%%%%%
\section{Mass hierarchies without fine-tuning}
\label{sec:mass_hierarchies}
%%%%%%%%%%%%%%%%%%%%%%%

%%%%%%%%%%%%%%%%%%%%%%%
\subsection{Mass matrices close to symmetric points}
\label{sec:theory}
%%%%%%%%%%%%%%%%%%%%%%%
%
In theories where modular invariance is broken only by the modulus, the flavour structure of mass matrices in the limit of unbroken supersymmetry is determined by the value of \(\tau\) and by the couplings in the superpotential.
At a symmetric point \(\tau = \tau_\text{sym}\), flavour textures can be severely constrained by the residual symmetry group, which may enforce the presence of multiple zero entries in the mass matrices.
As \(\tau\) moves away from its symmetric value, these entries will generically become non-zero. The magnitudes of such (residual-)symmetry-breaking entries will be controlled by the size of the departure \(\epsilon\) from \(\tau_\text{sym}\) and
by the field transformation properties under the residual symmetry group (which may depend on the modular weights). This is shown in what follows.

Consider a modular-invariant bilinear
\begin{equation}
  \label{eq:bilinear}
  \psi^c_i \, M(\tau)_{ij}\, \psi_j \,,
\end{equation}
where the superfields \(\psi\) and \(\psi^c\) transform under the modular group as%
\footnote{Note that in the case of a Dirac bilinear \(\psi\) and \(\psi^c\) are independent fields, so in general \(k^c \neq k\) and \(\rho^c \neq \rho, \rho^{*}\).}
\begin{equation}
\label{eq:psi_mod_trans}
\begin{split}
  \psi \,&\xrightarrow{\gamma}\, (c \tau + d)^{-k} \rho(\gamma) \,\psi \,, \\
  \psi^c \,&\xrightarrow{\gamma}\, (c \tau + d)^{-k^c} \rho^c(\gamma)\, \psi^c \,,
\end{split}
\end{equation}
so that each \(M(\tau)_{ij}\) is a modular form of level \(N\) and weight \(K \equiv k+k^c\).
Modular invariance requires \(M(\tau)\) to transform as
\begin{equation}
  \label{eq:mass_matrix_mod_trans}
  M(\tau)\, \xrightarrow{\gamma}\, M(\gamma \tau) = (c \tau + d)^K \rho^c(\gamma)^{*} M(\tau) \rho(\gamma)^{\dagger} \,.
\end{equation}
Taking \(\tau\) to be close to the symmetric point, and setting \(\gamma\) to the residual symmetry generator, one can use this transformation rule to constrain the form of the mass matrix \(M(\tau)\). We consider each of the three symmetric points in turn.

%%%%%%%%%%%%%%%%%%%%%%%
\subsubsection{\texorpdfstring{\(\tau_\text{sym} = i\infty\)}{tau sym = i infty}}
\label{sec:theory_T}
%%%%%%%%%%%%%%%%%%%%%%%
The representation basis for the group generators \(S\) and \(T\) typically found in the literature is the \(T\)-diagonal basis, in which \(\rho^{(c)}(T) = \diag (\rho^{(c)}_i)\). This basis is particularly useful for the analysis of models where \(\tau\) is `close' to \(\tau_\text{sym} = i\infty\), i.e.~models with large \(\im \tau\).
By setting \(\gamma = T\) in eq.~\eqref{eq:mass_matrix_mod_trans}, one finds
\begin{equation}
  \label{eq:mass_matrix_T_trans}
  M_{ij}(T \tau) = \left( \rho^c_i \rho_j \right)^{*} M_{ij}(\tau) \,.
\end{equation}
It is convenient to treat the \(M_{ij}\) as functions of \(q \equiv \exp \left( 2 \pi i \tau / N \right)\), so that 
\begin{equation}
  \label{eq:q}
  \epsilon \equiv |q| =  e^{- 2 \pi\im \tau / N}
\end{equation}
parameterises the deviation of \(\tau\) from the symmetric point.
Note that the entries \(M_{ij}(q)\) depend analytically on \(q\) and that \(q \xrightarrow{T} \zeta q\), with \(\zeta \equiv \exp \left( 2 \pi i / N \right)\). Thus, in terms of \(q\), eq.~\eqref{eq:mass_matrix_T_trans} reads
\begin{equation}
  M_{ij}(\zeta q) = (\rho^c_i \rho_j)^{*} M_{ij}(q) \,.
\end{equation}
Expanding both sides in powers of \(q\), one finds
\begin{equation}
  \zeta^n M_{ij}^{(n)}(0) = (\rho^c_i \rho_j)^{*} M_{ij}^{(n)}(0) \,,
\end{equation}
where \(M_{ij}^{(n)}\) denotes the \(n\)-th derivative of \(M_{ij}\) with respect to \(q\). This means that \(M_{ij}^{(n)}(0)\) can only be non-zero for values of \(n\) such that \((\rho^c_i \rho_j)^{*} = \zeta^n \).

It is clear that in the symmetric limit \(q \to 0\) the entry \(M_{ij} = M_{ij}^{(0)}\) is only allowed to be non-zero if \(\rho^c_i \rho_j = 1 \).
More generally, if \((\rho^c_i \rho_j)^{*} = \zeta^l\) with \(0 \leq l < N\),
\begin{equation}
  \label{eq:series_T}
  M_{ij}(q) = a_0\, q^l + a_1\, q^{N+l} + a_2\, q^{2N+l} + \ldots \,
\end{equation}
in the vicinity of the symmetric point.
It crucially follows that the entry \(M_{ij}\) is expected to be \(\mathcal{O}(\epsilon^l)\) whenever \(\im \tau\) is large. The power \(l\) only depends on how the representations of \(\psi\) and \(\psi^c\) decompose under the residual symmetry group \(\mathbb{Z}_N^T\).
This point will be made explicit in section~\ref{sec:decomp}.

%%%%%%%%%%%%%%%%%%%%%%%
\subsubsection{\texorpdfstring{\(\tau_\text{sym} = i\)}{tau sym = i}}
\label{sec:theory_S}
%%%%%%%%%%%%%%%%%%%%%%%
For the analysis of models where \(\tau\) is in the vicinity of \(\tau_\text{sym} = i\), it is convenient to switch to the basis where the \(S\) generator is represented by a diagonal matrix.
In this \(S\)-diagonal basis, one has \(\rho^{(c)}(S) =\diag (\rho^{(c)}_i)\).%
\footnote{Although we make use of the same notation, the \(\rho^{(c)}_i\) depend on the basis under consideration.}
It is useful to define and work with 
\begin{equation}
  \label{eq:rhotilde_S}
  \tilde\rho^{(c)}_i \equiv i^{k^{(c)}} \rho^{(c)}_i\,,
\end{equation}
which not only simplify the algebra, but also correspond to representations of the residual symmetry group, see eq.~\eqref{eq:phasefactors_S}.
By setting \(\gamma = S\) in eq.~\eqref{eq:mass_matrix_mod_trans}, one finds
\begin{equation}
  \label{eq:mass_matrix_S_trans}
  M_{ij}(S \tau) = (-i \tau)^K \left( \tilde\rho^c_i \tilde\rho_j \right)^{*} M_{ij}(\tau) \,.
\end{equation}
We now treat the \(M_{ij}\) as functions of
\begin{equation}
  s \equiv \frac{\tau - i}{\tau + i} \,, 
\end{equation}
so that, in this context, \(\epsilon \equiv |s|\) parameterises the deviation of \(\tau\) from the symmetric point.
Note that the entries \(M_{ij}(s)\) depend analytically on \(s\) and that \(s \xrightarrow{S} -s\). Thus, in terms of \(s\), eq.~\eqref{eq:mass_matrix_S_trans} reads
\begin{equation}
  M_{ij}(-s) =  \left( \frac{1+s}{1-s} \right)^K (\tilde\rho^c_i \tilde\rho_j)^{*} M_{ij}(s)
  \quad \Rightarrow \quad
  \tilde{M}_{ij}(-s) = (\tilde\rho^c_i \tilde\rho_j)^{*} \tilde{M}_{ij}(s) \,,
\end{equation}
where we have introduced \(\tilde{M}_{ij}(s) \equiv (1-s)^{-K} M_{ij}(s)\). Expanding both sides in powers of \(s\), one obtains
\begin{equation}
  \label{eq:expansion_S}
  (-1)^n \tilde{M}_{ij}^{(n)}(0) = (\tilde\rho^c_i \tilde\rho_j)^{*} \tilde{M}_{ij}^{(n)}(0) \,,
\end{equation}
where \(\tilde{M}_{ij}^{(n)}\) denotes the \(n\)-th derivative of \(\tilde{M}_{ij}\) with respect to \(s\).

It should be clear from eq.~\eqref{eq:expansion_S} that for \(\tau \simeq i\) the mass matrix entry \(M_{ij} \sim \tilde M_{ij}\) is only allowed to be \(\mathcal{O}(1)\) when \(\tilde\rho^c_i \tilde\rho_j = 1\).
If instead \(\tilde\rho^c_i \tilde\rho_j = -1\), the entry \(M_{ij} \sim \tilde M_{ij}\) is expected to be \(\mathcal{O}(\epsilon)\), with \(\epsilon = |s|\).
Note that, unlike in the previous section, the relevant factors \(\tilde \rho_i^{(c)}\) 
depend on the weights \(k^{(c)}\) via eq.~\eqref{eq:rhotilde_S}.

%%%%%%%%%%%%%%%%%%%%%%%
\subsubsection{\texorpdfstring{\(\tau_\text{sym} = \omega\)}{tau sym = omega}}
\label{sec:theory_ST}
%%%%%%%%%%%%%%%%%%%%%%%
Finally, for the analysis of models where \(\tau\) is in the vicinity of \(\tau_\text{sym} = \omega\), we consider the basis where the product \(ST\) is represented by a diagonal matrix.
In this \(ST\)-diagonal basis where \(\rho^{(c)}(ST) =\diag (\rho^{(c)}_i)\), it is useful to define
\begin{equation}
  \label{eq:rhotilde_ST}
  \tilde\rho^{(c)}_i \equiv \omega^{k^{(c)}} \rho^{(c)}_i\,,
\end{equation}
which are representations under the residual symmetry group, see eq.~\eqref{eq:phasefactors_ST}.
By setting \(\gamma = ST\) in eq.~\eqref{eq:mass_matrix_mod_trans}, one finds
\begin{equation}
  \label{eq:mass_matrix_ST_trans}
  M_{ij}(ST \tau) = [-\omega (\tau + 1)]^K \left( \tilde\rho^c_i \tilde\rho_j \right)^{*} M_{ij}(\tau) \,.
\end{equation}
It is now convenient to treat the \(M_{ij}\) as functions of
\begin{equation}
  u \equiv \frac{\tau - \omega}{\tau - \omega^2} \,, 
\end{equation}
so that, in this context, \(\epsilon \equiv |u|\) parameterises the deviation of \(\tau\) from the symmetric point. 
Note that the entries \(M_{ij}(u)\) depend analytically on \(u\) and that \(u \xrightarrow{ST} \omega^2 u\). Thus, in terms of \(u\), eq.~\eqref{eq:mass_matrix_ST_trans} reads
\begin{equation}
  M_{ij}(\omega^2 u) =  \left( \frac{1-\omega^2 u}{1-u} \right)^K
  (\tilde\rho^c_i \tilde\rho_j)^{*} M_{ij}(u)
  \quad \Rightarrow \quad
  \tilde{M}_{ij}(\omega^2 u) = (\tilde\rho^c_i \tilde\rho_j)^{*} \tilde{M}_{ij}(u) \,,
\end{equation}
where \(\tilde{M}_{ij}(u) \equiv (1-u)^{-K} M_{ij}(u)\). Expanding both sides in powers of \(u\), one obtains
\begin{equation}
  \label{eq:expansion_ST}
  \omega^{2n} \tilde{M}_{ij}^{(n)}(0) = (\tilde\rho^c_i \tilde\rho_j)^{*} \tilde{M}_{ij}^{(n)}(0) \,,
\end{equation}
where \(\tilde{M}_{ij}^{(n)}\) denotes the \(n\)-th derivative of \(\tilde{M}_{ij}\) with respect to \(u\).

It follows that for \(\tau \simeq \omega\) the mass matrix entry \(M_{ij} \sim \tilde M_{ij}\) is only allowed to be \(\mathcal{O}(1)\) when \(\tilde\rho^c_i \tilde\rho_j = 1\). More generally, if \(\tilde\rho^c_i \tilde\rho_j = \omega^l\) with \(l=0,1,2\), then the entry \(M_{ij} \sim \tilde M_{ij}\) is expected to be \(\mathcal{O}(\epsilon^l)\) in the vicinity of \(\tau = \omega\), with \(\epsilon = |u|\).
Like in the previous section, the factors \(\tilde \rho_i^{(c)}\) depend on the weights \(k^{(c)}\), see eq.~\eqref{eq:rhotilde_ST}.

%%%%%%%%%%%%%%%%%%%%%%%
\subsection{Decomposition under residual symmetries}
\label{sec:decomp}
%%%%%%%%%%%%%%%%%%%%%%%
%
We have just shown that, as \(\tau\) departs from a symmetric value \(\tau_\text{sym}\), the entries of fermion mass matrices are of \(\mathcal{O}(\epsilon^l)\), where \(\epsilon\) parameterises the deviation of \(\tau\) from \(\tau_\text{sym}\).
The powers \(l\) are extracted from products of factors which, in this section, are shown to correspond to representations of the residual symmetry group.
One can systematically identify these residual symmetry representations for the different possible choices of \(\Gamma_N'\) representations of matter fields.
This knowledge will later be exploited to construct hierarchical mass matrices via controlled corrections to entries which are zero in the symmetric limit. 

We start by noting that matter fields \(\psi\) furnish `weighted' representations \((\mathbf{r}, k)\) of the finite modular group \(\Gamma_N'\). Whenever a residual symmetry is preserved by the value of \(\tau\), matter fields decompose into unitary representations of the residual symmetry group.
Modulo a possible \(\mathbb{Z}_2^R\) factor,%
\footnote{See the discussion in appendix~\ref{app:decomp}.}
these groups are the cyclic groups \(\mathbb{Z}_N^T\),  \(\mathbb{Z}_4^S\), and  \(\mathbb{Z}_3^{ST}\) (cf.~section~\ref{sec:residual}).
A cyclic group \(\mathbb{Z}_n \equiv \left\langle a \, \vert \, a^n = 1 \right\rangle\) has \(n\) inequivalent 1-dimensional irreducible representations (irreps) \(\mathbf{1}_k\), where \(k = 0, \ldots, n-1\) is sometimes referred to as a charge. The group generator \(a\) is represented by one of the \(n\)-th roots of unity,
\begin{equation}
  \label{eq:Zn_reps}
  \mathbf{1}_k \, : \quad \rho(a) = \exp \left( 2\pi i \frac{k}{n} \right) \, .
\end{equation}
For odd \(n\), the only real irrep of \(\mathbb{Z}_n\) is the trivial one, \(1_0\) (the reality of an irrep is indicated by removing the boldface). For even \(n\), there is one more real irrep, \(1_{n/2}\). All other irreps are complex, and split into pairs of conjugated irreps: \((\mathbf{1}_k)^{*} = \mathbf{1}_{n-k}\).

To illustrate the aforementioned decomposition of representations at symmetric points, we take as an example a \((\mathbf{3}, k)\) triplet \(\psi\) of \(S_4'\). It transforms under the unbroken \(\gamma = ST\) at \(\tau = \omega\) as
\begin{equation}
  \psi_i \,\xrightarrow{ST}\, \left( -\omega - 1 \right)^{-k} \rho_{\mathbf{3}}(ST)_{ij} \,\psi_j
  = \omega^k \rho_{\mathbf{3}}(ST)_{ij}\, \psi_j \,.
\end{equation}
One can check that the eigenvalues of \(\rho_{\mathbf{3}}(ST) \) are \(1\), \(\omega\) and \(\omega^2\), and so in a suitable (\(ST\)-diagonal) basis the transformation rule explicitly reads
\begin{equation}
  \label{eq:explicitST}
  \psi \,\xrightarrow{ST}\, \omega^k
  \begin{pmatrix}
    1 & 0 & 0 \\
    0 & \omega & 0 \\
    0 & 0 & \omega^2
  \end{pmatrix}
  \psi 
  =
  \begin{pmatrix}
    \omega^k & 0 & 0 \\
    0 & \omega^{k+1} & 0 \\
    0 & 0 & \omega^{k+2}
  \end{pmatrix}
  \psi \,,
\end{equation}
which means that \(\psi\) decomposes as \(\psi \leadsto \mathbf{1}_k \oplus \mathbf{1}_{k+1} \oplus \mathbf{1}_{k+2}\) under the residual \(\mathbb{Z}_3^{ST}\).

One can find the residual symmetry representations for any other `weighted' multiplet of a finite modular group in a similar fashion. For a given level \(N\), the decompositions of fields under a certain residual symmetry group only depend on the pair \((\mathbf{r},k)\). In general:
\begin{itemize}
    \item At \(\tau = i\infty\), \(\psi \sim (\mathbf{r}, k)\) transforms under the unbroken \(\gamma = T\) as
    \begin{equation}
    \psi_i \,\xrightarrow{T}\,  \rho_{\mathbf{r}}(T)_{ij} \,\psi_j
  = \rho_i\, \psi_i \,,
    \end{equation}
    where for the last equality we have assumed to be in a \(T\)-diagonal basis (no sum over \(i\)). The phase factors \(\rho_i\) correspond to the \(\mathbb{Z}_N^T\) irreps into which \(\psi\) decomposes. It follows that each \(\rho_i\) is a power of \(\zeta = \exp(2\pi i/N)\), depending on \(\mathbf{r}\) but not on \(k\).
    \item At \(\tau = i\), \(\psi \sim (\mathbf{r}, k)\) transforms under the unbroken \(\gamma = S\) as
    \begin{equation}
    \label{eq:phasefactors_S}
    \psi_i \,\xrightarrow{S}\,  (-i)^{-k}\rho_{\mathbf{r}}(S)_{ij} \,\psi_j
  = i^k \rho_i\, \psi_i \,,
    \end{equation}
    where for the last equality we have assumed to be in an \(S\)-diagonal basis (no sum over \(i\)). The phase factors \(\tilde\rho_i = i^k \rho_i\) correspond to the \(\mathbb{Z}_4^S\) irreps into which \(\psi\) decomposes. It follows that each \(\tilde \rho_i\) is a power of \(i\) which depends both on \(\mathbf{r}\) and on \(k\,(\text{mod }4)\).
    \item At \(\tau = \omega\), \(\psi \sim (\mathbf{r}, k)\) transforms under the unbroken \(\gamma = ST\) as
    \begin{equation}
    \label{eq:phasefactors_ST}
    \psi_i \,\xrightarrow{ST}\,  (-\omega-1)^{-k}\rho_{\mathbf{r}}(ST)_{ij} \,\psi_j
  = \omega^k \rho_i\, \psi_i \,,
    \end{equation}
    where for the last equality we have assumed to be in an \(ST\)-diagonal basis (no sum over \(i\)), as in the explicit example of eq.~\eqref{eq:explicitST}. The phase factors \(\tilde\rho_i = \omega^k \rho_i\) correspond to the \(\mathbb{Z}_3^{ST}\) irreps into which \(\psi\) decomposes. It follows that each \(\tilde \rho_i\) is a power of \(\omega\) which depends both on \(\mathbf{r}\) and on \(k\,(\text{mod }3)\).
\end{itemize}

After identifying the \(\tildepar{\rho}_i\) and \(\tildepar{\rho}_i^{\,\,c}\) factors for the fields \(\psi\) and \(\psi^c\) entering a bilinear (equivalently, their irrep decompositions), one can apply the results of the previous section to determine the structure of a mass matrix in the vicinity of a symmetric point in terms of powers of \(\epsilon\), in the appropriate basis.
It follows from the above that, in the analysis with large \(\im \tau\), the product \((\rho^c_i \rho_j)^{*}\) matches some power \( \zeta^l\) with \(0 \leq l < N\), while in the analysis corresponding to \(\tau \simeq \omega\) one necessarily has \(\tilde\rho^c_i \tilde\rho_j = \omega^l\) with \(l=0,1,2\). These were tacitly taken as the most general possibilities in sections~\ref{sec:theory_T} and~\ref{sec:theory_ST}.
The same reasoning implies that, in the \(\tau \simeq i\) context, \(\tilde\rho^c_i \tilde\rho_j\) is some integer power \(i^l\), with \(l=0,1,2,3\). It turns out that only two out of the four possibilities are viable, namely \(l = 0, 2\) so that \(\tilde\rho^c_i \tilde\rho_j = \pm 1\), as considered in section~\ref{sec:theory_S}. This is due to the fact that \(M(\tau)_{ij}\) is \(R\)-even and thus the fields \(\psi_i^c\) and \(\psi_j\) need to carry the same \(R\)-parity (see also appendix~\ref{app:decomp}).

We list in Tables~\ref{tab:S3_residual_reps}\,--\,\ref{tab:A5p_residual_reps} of appendix~\ref{app:decomp} the decompositions of the weighted representations of \(\Gamma'_N\) (\(N \leq 5\)) under the three residual symmetry groups, i.e.~the residual decompositions of the irreps of \(\Gamma_2' \simeq S_3\), \(\Gamma_3' \simeq A_4' = T'\), \(\Gamma_4' \simeq S_4' = SL(2,\mathbb{Z}_4)\), and \(\Gamma_5' \simeq A_5' = SL(2,\mathbb{Z}_5)\).

%%%%%%%%%%%%%%%%%%%%%%%
\subsection{Hierarchical structures}
\label{sec:hierarchies}
%%%%%%%%%%%%%%%%%%%%%%%
%
%%%%%%%%%%%%%%%%%%%%%%%
\subsubsection{From entries to masses}
\label{sec:masses}
%%%%%%%%%%%%%%%%%%%%%%%
We are in a position to use the results found so far and construct hierarchical mass matrices in the vicinity of a symmetric point. We have seen that in an appropriate basis \(M(\tau(\epsilon))_{ij} \sim \mathcal{O}(\epsilon^{l})\). For each \((i,j)\) pair, the power \(l\) can be obtained from the residual symmetry group decompositions of Tables~\ref{tab:S3_residual_reps}\,--\,\ref{tab:A5p_residual_reps}.

Note that a modular-symmetric mass matrix \(M(\tau(\epsilon))\) depends analytically on the small real parameter \(\epsilon\), defined in section~\ref{sec:theory} for each symmetric point. Physical masses are the singular values of \(M(\tau)\) and are also analytic functions of \(\epsilon\).%
\footnote{More precisely, the elements of the unordered tuple of non-zero singular values are absolute values of analytic functions of \(\epsilon\), see Theorem 4.3.17 in Ref.~\cite{Hinrichsen2005}.}
After the modular symmetry breaking, the leading superpotential contribution to each fermion mass is thus expected to be proportional to a power of \(\epsilon\) which depends on the hierarchical structure of the entries of \(M\).
To find out which, one can make use of the following set of relations, valid for any \(n\times n\)  complex matrix \(M\)~\cite{Marzocca:2014tga}:
\begin{equation}
  \sum_{i_1 < \ldots < i_p}m_{i_1}^2\ldots m_{i_p}^2 =
  \sum \left| \det M_{p\times p}\right|^2
  \,,
\end{equation}
where \(p=1,\ldots,n\) is fixed, \(m_i\) are the singular values of \(M\), and the sum on the right-hand side goes over all possible \(p\times p\) submatrices \(M_{p\times p}\) of \(M\).
In the particular case of \(n=3\), we denote the masses by \((m_1,m_2,m_3)\) such that their leading terms are respectively of order \((\epsilon^{d_1},\epsilon^{d_2},\epsilon^{d_3})\) with \(d_1 \geq d_2 \geq d_3 \geq 0\). Then, 
\begin{equation}
  \label{eq:power_counting}
\begin{split}
  m_3^2 
  &\,\sim\, \sum_{i,j} |M_{ij}|^2 = \tr M^\dagger M\,,
  \\[-1mm]
  m_2^2 m_3^2 
  &\,\sim\, \sum |\det M_{2 \times 2}|^2
  \,\,\,\Rightarrow\,\,\,\,
  m_2^2 \,\sim\, \frac{\sum |\det M_{2 \times 2}|^2}{\tr M^\dagger M}
  \,,
  \\
  m_1^2 m_2^2 m_3^2 
  &\,=\, \left|\det M \right|^2
  \qquad\quad\,\;\,\Rightarrow\,\,\,\,
  m_1^2 \,\sim \, \frac{\left|\det M \right|^2}{\sum |\det M_{2 \times 2}|^2}
  \,,
\end{split}
\end{equation}
where \(\sim\) refers to power counting in \(\epsilon\) and not necessarily to a reliable approximation. Note that so far the considered mass spectrum is generic. This is to be contrasted with the special case of a {\it hierarchical} \(3 \times 3\) mass matrix, for which \(d_1 > d_2 > d_3 \geq 0\) and thus \(m_1 \ll m_2 \ll m_3\). In this case, eqs.~\eqref{eq:power_counting} turn into useful approximations,
\begin{equation}
\begin{split}
  m_3^2
  &\,\simeq\, \sum_i m_i^2 = \tr M^\dagger M\,,
  \\
  m_2^2 m_3^2 
  &\,\simeq\, \sum_{i<j} m_i^2 m_j^2 = \frac{1}{2} \left((\tr M^\dagger M)^2 - \tr (M^\dagger M)^2 \right)
  \,,
\end{split}
\end{equation}
and lead to reliable expressions for \(m_3\), \(m_2\) and \(m_1 = |\det M|/(m_2m_3)\).

%%%%%%%%%%%%%%%%%%%%%%%
\subsubsection{Example and results}
\label{sec:example_results}
%%%%%%%%%%%%%%%%%%%%%%%
As an example of application of our results, consider a model at level \(N=5\) with \(\tau\) having a large imaginary part and with matter fields in weighted representations \(\psi \sim (\mathbf{3},k)\) and \(\psi^c \sim (\mathbf{3}',k^c)\). From Table~\ref{tab:A5p_residual_reps} one sees that \(\psi \leadsto {1}_0 \oplus \mathbf{1}_{1} \oplus \mathbf{1}_{4}\)  and \(\psi^c \leadsto {1}_0 \oplus \mathbf{1}_{2} \oplus \mathbf{1}_{3}\)  under the residual group at the symmetric point \(\tau_\text{sym} = i \infty\). One can then identify \(\rho_i = \diag(1, \zeta, \zeta^4)\) and \(\rho_i^c = \diag(1,\zeta^2, \zeta^3)\), with \(\zeta = \exp(2\pi i/5)\), which allows for the structure
\begin{equation}
  \label{eq:example}
  M(\tau(\epsilon)) \sim 
  \begin{pmatrix}
  1 & \epsilon^4 & \epsilon \\
  \epsilon^3 & \epsilon^2 & \epsilon^4  \\
  \epsilon^2  & \epsilon & \epsilon^3 
  \end{pmatrix}\,,
  \quad \text{with }\epsilon =  e^{- 2 \pi\im \tau / 5}\,.
\end{equation}
Resorting to~\eqref{eq:power_counting}, one finds that the spectrum is hierarchical, with \((m_3,m_2,m_1) \,\sim\, (1,\epsilon,\epsilon^4)\).

Note that to have a non-zero mass matrix one needs the sum \(K=k+k^c\) to be even (in this case), since matter fields furnish unhatted representations of the finite modular group and should carry the same \(R\)-parity (see appendix~\ref{app:decomp}).
Furthermore, in order to obtain the full structure of eq.~\eqref{eq:example} and the expected hierarchical spectrum, \(K\) must be large enough that sufficient modular forms contribute to \(M(\tau)\). For instance, for \(K=2\) the superpotential may turn out to include a unique contribution:
\begin{equation}
  W \,\supset\, \sum_s \alpha_s \left(Y_\mathbf{5}^{(5,2)}(\tau) \psi^c \psi\right)_{\mathbf{1},s}
  \quad\Rightarrow\quad M(\tau) = \alpha
  \begin{pmatrix}
    \sqrt{3} Y_1 & Y_5 & Y_2 \\
    Y_4 & -\sqrt{2} Y_3 & -\sqrt{2} Y_5 \\
    Y_3 & -\sqrt{2} Y_2 & -\sqrt{2} Y_4
  \end{pmatrix}_{Y_\mathbf{5}^{(5,2)}}\,,
\end{equation}
where the \(\alpha_s\) are coupling constants, the sum is taken over all possible singlets \(s\) and \(Y_\mathbf{r(,\mu)}^{(N,K)}\) denotes the modular form multiplet of level \(N\), weight \(K\) and irrep \(\mathbf{r}\), with \(\mu\) possibly labelling linearly independent multiplets of the same type. The rightmost matrix subscript indicates the multiplet to which the \(Y_i\) components belong. We have considered the \(T\)-diagonal basis for \(A_5'\). One can explicitly check that, at leading order in \(\epsilon = |q|\), the components of  \(Y_{\mathbf{5}}^{(5,2)}(\tau)\) read \(\left( Y_1, Y_2, Y_3, Y_4, Y_5\right) \simeq \mathcal{N}\left( -1/\sqrt{6},\, q,\, 3 q^2,\, 4 q^3,\, 7 q^4 \right)\), with \(q = \exp \left( 2 \pi i \tau / 5 \right)\) and a common normalisation \(\mathcal{N}\).
The power structure matches that of eq.~\eqref{eq:example} and naïvely this corresponds to the desired \((1,\epsilon,\epsilon^4)\) spectrum. Upon closer inspection, however, one realises that the determinant of \(M\) vanishes identically for {\it any} value of \(\tau\),
\begin{equation}
\det M \,\propto\, 
\sqrt{6}\, Y_1 Y_3 Y_4 -Y_2^2 Y_4 + Y_2 \big(Y_3^2 - \sqrt{6}\, Y_1 Y_5\big) + Y_5 \big(Y_4^2 - Y_3 Y_5\big)
= 0\,,
\end{equation}
meaning that at least one fermion is massless. In the vicinity of \(\tau_\text{sym} = i \infty\), we have \((m_3,m_2,m_1) \,\sim\, (1,\epsilon,0)\). This issue is resolved already at weight \(K=4\), for which the modular multiplets \(Y_\mathbf{4}^{(5,4)}\), \(Y_{\mathbf{5},1}^{(5,4)}\), and \(Y_{\mathbf{5},2}^{(5,4)}\) are available. In this case the spectrum follows a \((1,\epsilon,\epsilon^4)\) pattern, without a massless fermion (see section~\ref{sec:models_A5_3x3'}).

\vskip 2mm

Let us pause and describe our philosophy going forward. We are interested in identifying  \(3\times 3\) hierarchical mass matrices where the hierarchical pattern is a result of the proximity of the modulus to a point of residual symmetry and no massless fermions are present in the spectrum. We assume to effectively be dealing with bilinears of the type~\eqref{eq:bilinear} and consider all possible 3-dimensional representations for the fields \(\psi\) and \(\psi^c\). While the representations \(\mathbf{r}\) and \(\mathbf{r}^c\) are in general reducible, we focus on the case where the same weight is shared between the irreps into which they decompose.%
\footnote{
The freedoms in choosing i) the normalisations of modular multiplets and ii) the normalisations of Clebsch-Gordan coefficients introduce ambiguities in the identification of hierarchies. 
In the interest of minimizing their impact, when possible we make use of modular form multiplets obtained from tensor products of a single \(k=1\) multiplet with itself, via canonically normalised Clebsch-Gordan coefficients, as in~\cite{Novichkov:2020eep}.
Using this procedure, one expects that relative normalisations of modular multiplets cannot be responsible for hierarchies, at least within the same weight \(k\).
}
Thus, in our search, we take \(\mathbf{r}^{(c)}\) to be either irreducible or a direct sum of irreps sharing the same \(\rho(R)\). While it is possible for \(\mathbf{r}^{(c)}\) to be a direct sum of hatted and unhatted representations, the requirement of a common weight \(k^{(c)}\) would result in the co-existence of \(R\)-odd and \(R\)-even fields within \(\psi^{(c)}\). The fact that \(M(\tau)\) is \(R\)-even would then imply the isolation of these sectors by the \(\mathbb{Z}_2^R\) symmetry and the vanishing of some mixing angles.

Finally, it is straightforward to recognise that if all mass matrix entries are either \(\mathcal{O}(1)\) or \(\mathcal{O}(\epsilon)\), then leading contributions to the masses themselves are not expected to be smaller than \(\mathcal{O}(\epsilon)\), unless one resorts to cancellations. Therefore, for \(\tau \simeq i\) one cannot produce the desired hierarchical patterns solely as a consequence of the smallness of \(\epsilon\).

\vskip 2mm

The result of our analysis is given in Tables~\ref{tab:S3_patterns}\,--\,\ref{tab:A5p_patterns} of appendix~\ref{app:patterns}. These tables summarise, for each of the levels \(N \leq 5\), the patterns which may arise in the vicinity of the two potentially viable symmetric points, \(\tau_\text{sym} = \omega\) and \(i \infty\), for all \((\mathbf{r},\mathbf{r}^c)\) pairs of 3-dimensional representations and all weights \(k^{(c)}\).
One finds that it is only possible to obtain {\it hierarchical} spectra for a small list of representation pairs, the most promising of which are collected here, in Table~\ref{tab:good_patterns}.
%
%%%%%%%%%%%%%%%%%%%%%%%
\begin{table}[ht!]
\small
  \centering
  \renewcommand{\arraystretch}{1.5}
  \begin{tabular}{cccll}
\toprule
\belowrulesepcolor{light-gray}
\rowcolor{light-gray}
\(\,N\,\) & \(\,\Gamma'_N\,\) & \(\,\)Pattern\(\,\) & Sym.~point & Viable \(\mathbf{r} \otimes \mathbf{r}^c\quad\) \\
\aboverulesepcolor{light-gray}
\midrule
%%%%%%%%%%%%%%
  2 
& \(S_3\) 
& \((1,\epsilon,\epsilon^2)\)                
& \(\tau \simeq \omega\)  
& \([\mathbf{2}\oplus\mathbf{1}^{(\prime)}] \otimes [\mathbf{1}\oplus\mathbf{1}^{(\prime)}\oplus\mathbf{1}']\)
\\[2mm]
%%%%%%%%%%%%%%
\rowcolor{lighter-gray}
& &
& \(\tau \simeq \omega\)
& \([\mathbf{1}_a\oplus\mathbf{1}_a\oplus\mathbf{1}_a'] \otimes [\mathbf{1}_b\oplus\mathbf{1}_b\oplus\mathbf{1}_b'']\)
\\[1mm]
\rowcolor{lighter-gray}
\multirow{-2}{*}{3}
& \multirow{-2}{*}{\(A_4'\) }
& \multirow{-2}{*}{\((1,\epsilon,\epsilon^2)\)}
& \(\tau \simeq i \infty\)
& \([\mathbf{1}_a\oplus\mathbf{1}_a\oplus\mathbf{1}_a'] \otimes [\mathbf{1}_b\oplus\mathbf{1}_b\oplus\mathbf{1}_b'']\) with~\(\mathbf{1}_a \neq (\mathbf{1}_b)^*\)
\\[2mm]
%%%%%%%%%%%%%%
\multirow{3}{*}{4}
& \multirow{3}{*}{\(S_4'\)}
& \((1,\epsilon,\epsilon^2)\) 
& \(\tau \simeq \omega\)
& \([
\mathbf{3}_a \text{, or }
\mathbf{2}\oplus\mathbf{1}^{(\prime)} \text{, or }
\mathbf{\hat{2}}\oplus\mathbf{\hat{1}}^{(\prime)}
] \otimes [\mathbf{1}_b\oplus\mathbf{1}_b\oplus\mathbf{1}_b']\)
\\[1mm]
&
& \multirow{2}{*}{\((1,\epsilon,\epsilon^3)\)}
& \multirow{2}{*}{\(\tau \simeq i \infty\)}
& 
\(\mathbf{3}\hphantom{'} \otimes [\mathbf{2}\oplus\mathbf{1} \text{, or }
\mathbf{1}\oplus\mathbf{1}\oplus\mathbf{1}']\), 
\(\mathbf{3}' \otimes [\mathbf{2}\oplus\mathbf{1}' \text{, or }
\mathbf{1}\oplus\mathbf{1}'\oplus\mathbf{1}']\), 
\\[-1mm]
& & & & 
\(\mathbf{\hat{3}}' \otimes [\mathbf{\hat{2}}\oplus\mathbf{\hat{1}} \text{, or }
\mathbf{\hat{1}}\oplus\mathbf{\hat{1}}\oplus\mathbf{\hat{1}}']\),
\(\mathbf{\hat{3}}\hphantom{'} \otimes [\mathbf{\hat{2}}\oplus\mathbf{\hat{1}'} \text{, or }
\mathbf{\hat{1}}\oplus\mathbf{\hat{1}}'\oplus\mathbf{\hat{1}}']\)
\\[2mm]
%%%%%%%%%%%%%%
\rowcolor{lighter-gray}
5
& \(A_5'\)
& \((1,\epsilon,\epsilon^4)\) 
& \(\tau \simeq i \infty\)
& \(\mathbf{3}\otimes\mathbf{3}'\)  
\\
\aboverulesepcolor{lighter-gray}
\bottomrule
  \end{tabular}
  \caption{
      Hierarchical mass patterns which can be realised in the vicinity of symmetric points. These patterns are unaffected by the exchange \(\mathbf{r}\leftrightarrow \mathbf{r}^c\) and may only be viable for certain weights (see appendix~\ref{app:patterns}). Subscripts run over irreps of a certain dimension, and \(\mathbf{1}_a'''=\mathbf{1}_a\) for \(N=3\), while \(\mathbf{1}_a''=\mathbf{1}_a\) for \(N=4\). Primes in parenthesis are uncorrelated.
  }
  \label{tab:good_patterns}
\end{table}
%%%%%%%%%%%%%%%%%%%%%%%
%

\noindent
We have excluded from this summary table reducible representations made up of three copies of the same singlet, since in those cases at least three independent modular multiplets of the same type must contribute to the mass matrix to avoid a massless fermion, and the number of superpotential parameters is unappealingly high. Still, such cases may result in other interesting hierarchical patterns such as \((1,\epsilon^2,\epsilon^3)\) and \((\epsilon,\epsilon^2,\epsilon^3)\) and can be found in the tables of appendix~\ref{app:patterns}.

\vskip 2mm

Inspired by these results, we now turn to the construction of realistic models of lepton flavour where mass hierarchies are a consequence of the described mechanism.

%%%%%%%%%%%%%%%%%%%%%%%
\subsection{Promising models}
\label{sec:models_inf}
%%%%%%%%%%%%%%%%%%%%%%%
%
To ascertain the viability of modular-invariant models of lepton flavour one must confront their predictions with experimental data on ratios of charged-lepton masses, neutrino mass-squared differences and leptonic mixing angles, see Table~\ref{tab:globalFit} (the constraints on the Dirac CPV phase \(\delta\) are ignored in our fit). The reader is referred to~\cite{Novichkov:2018ovf} for details on our numerical procedure.

%%%%%%%%%%%%%%%%%%%%%%
\begin{table}[t]
\centering
\renewcommand{\arraystretch}{1.2}
\begin{tabular}{l|cc} 
\toprule
Observable & \multicolumn{2}{c}{Best-fit value and \(1\sigma\) range} \\ 
\midrule
\(m_e / m_\mu\) & \multicolumn{2}{c}{\(0.0048 \pm 0.0002\)} \\
\(m_\mu / m_\tau\) & \multicolumn{2}{c}{\(0.0565 \pm 0.0045\)} \\ 
\midrule
& NO & IO \\
\(\delta m^2/(10^{-5}\text{ eV}^2)\) & \multicolumn{2}{c}{\(7.34^{+0.17}_{-0.14}\)} \\
\(|\Delta m^2|/(10^{-3}\text{ eV}^2)\) & \(2.485^{+0.029}_{-0.032}\) & \(2.465^{+0.030}_{-0.031}\) \\
\(r \equiv \delta m^2/|\Delta m^2|\) & \(0.0295\pm0.0008\) & \(0.0298\pm0.0008\)\\
\(\sin^2\theta_{12}\) & \(0.305^{+0.014}_{-0.013}\) & \(0.303^{+0.014}_{-0.013}\) \\
\(\sin^2\theta_{13}\) & \(0.0222^{+0.0006}_{-0.0008}\) & \(0.0223^{+0.0007}_{-0.0006}\) \\
\(\sin^2\theta_{23}\) & \(0.545^{+0.020}_{-0.047}\) & \(0.551^{+0.016}_{-0.034}\) \\
\(\delta/\pi\) & \(1.28^{+0.38}_{-0.18}\)  & \(1.52^{+0.13}_{-0.15}\) \\
\bottomrule
\end{tabular}
\caption{Best-fit values and 1\(\sigma\) ranges for 
neutrino oscillation parameters obtained from the global analysis~\cite{Capozzi:2020qhw}, and for charged-lepton mass ratios,
given at the scale \(2\times 10^{16}\) GeV with the \(\tan \beta\) averaging
described in~\cite{Feruglio:2017spp}, obtained from Ref.~\cite{Ross:2007az}.
The parameters entering the definition of \(r\) are \(\delta m^2 \equiv m_2^2-m_1^2\)
and \(\Delta m^2 \equiv m_3^2 - (m_1^2+m_2^2)/2\).}
\label{tab:globalFit}
\end{table}
%%%%%%%%%%%%%%%%%%%%%%
%

The minimal-form Kähler potential is here considered,
\begin{equation}
K(\tau, \overline{\tau}, \psi_I, \overline{\psi}_I)
= - \Lambda_0^2 \log(-i\tau + i \overline{\tau})
  + \sum_I \frac{|\psi_I|^2}{(-i\tau + i \overline{\tau})^{k_I}} \,, 
\label{eq:Kahler}
\end{equation}
with \(\Lambda_0\) of mass dimension one. 
We further take Higgs doublets \(H_u\) and \(H_d\) to be singlets under the modular group. Charged-lepton masses are obtained from their Yukawa interactions,
\begin{equation}
 W \supset \sum_s \alpha_s \left(Y_{\mathbf{r}_s}^{(N,k_Y)}(\tau)\, E^c \,L \right)_{\mathbf{1},s} \!H_d\,,
\end{equation}
where \(L\) and \(E^c\) denote lepton doublet and charged-lepton singlet superfields with weights \(k_L\) and \(k_E\), respectively. Neutrino masses are generated either by the Weinberg operator,
\begin{equation}
 W \supset   
 \frac{1}{\Lambda} \sum_s g_s \left(Y_{\mathbf{r}_s}^{(N,k_W)}(\tau) \,L^2 \right)_{\mathbf{1},s}\!H_u^2
 \,,
\end{equation}
or within a type-I seesaw UV completion,
\begin{equation}
 W \supset  \sum_s g_s \left(Y_{\mathbf{r}_s}^{(N,k_\mathcal{Y})}(\tau) \,N^c \,L \right)_{\mathbf{1},s} \!H_u
 \,+\,  \sum_s \Lambda_s \left(Y_{\mathbf{r}_s}^{(N,k_M)}(\tau) \,(N^c)^2
 \right)_{\mathbf{1},s} \,,
\end{equation}
where at least 2 neutrino gauge-singlet superfields \(N^c\) of weight \(k_N\) are present in the model.  
To compensate the modular weights of field monomials, the modular forms entering the Weinberg and Majorana terms need to have weights \(k_W = 2k_L\) and \(k_M = 2k_N\), while those in Yukawa terms need instead \(k_Y = k_L + k_E\) and \(k_\mathcal{Y} = k_L + k_N\).

The relevant superpotentials can be cast in the form
\begin{equation}
    W \,= \,\lambda_{ij} \,E^c_i\, L_j\, H_d + \begin{cases}
 \dfrac{1}{2} \,c_{ij}\, L_i \,L_j \,H_u^2 &\,\,\text{(Weinberg)}\\[2mm]
 \mathcal{Y}_{ij} \,N^c_i\, L_j\, H_u + \dfrac{1}{2}\, (M_N)_{ij} \,N^c_i\, N^c_j &\,\,\text{(Seesaw)}
\end{cases}\,\,.
\end{equation}
After electroweak symmetry breaking, with \(\langle H_u\rangle = (0,v_u)^T\) and \(\langle H_d\rangle = (v_d,0)^T\), these result in the Lagrangian mass terms for leptons
\begin{equation}
  \mathcal{L} \,\supset\,
  -\big(M_e\big)_{ij}\,\overline{e_{iL}}\,e_{jR}
  -\frac{1}{2}\,\big(M_\nu\big)_{ij}\,\overline{\nu_{iR}^c}\,\nu_{jL}
  + \text{h.c.}\,,
\end{equation}
which have been written in terms of four-spinors. Here, \(M_e = v_d \lambda^\dagger\), while
\begin{equation}
M_\nu \,=\, \begin{cases}
 v_u^2\, c &\,\,\text{(Weinberg)}\\[2mm]
 -v_u^2\,\mathcal{Y}^T\,M_N^{-1}\,\mathcal{Y} &\,\,\text{(Seesaw)}
\end{cases}\,\,.
\end{equation}
Finally, aiming at minimal and predictive examples, we impose a generalised CP symmetry on the models, enforcing the reality of coupling constants~\cite{Novichkov:2019sqv}.

%%%%%%%%%%%%%%%%%%%%%%%
\subsubsection{\texorpdfstring{\(A_5'\)}{A5'} models with \texorpdfstring{\(L \sim \mathbf{3}\), \(E^c \sim \mathbf{3}'\)}{L=3, Ec=3'}}
\label{sec:models_A5_3x3'}
%%%%%%%%%%%%%%%%%%%%%%%
%

We start by considering the most `structured' series of hierarchical models, i.e.~the case with both fields \(L\), \(E^c\) furnishing complete irreducible representations of the finite modular group.
According to Table~\ref{tab:good_patterns}, the only such possibility arises at level~\(N = 5\) in the vicinity of~\(\tau = i \infty\) when \(L\) and \(E^c\) are different triplets of the finite modular group \(A_5'\).
For definiteness, we choose \(L \sim (\mathbf{3}, k_L)\), \(E^c \sim (\mathbf{3'}, k_E)\).
The predicted charged-lepton mass pattern is \((m_{\tau}, m_{\mu}, m_e) \sim (1, \epsilon, \epsilon^4)\).

\begin{figure}[t!]
  \centering
  \includegraphics[width=0.65\textwidth]{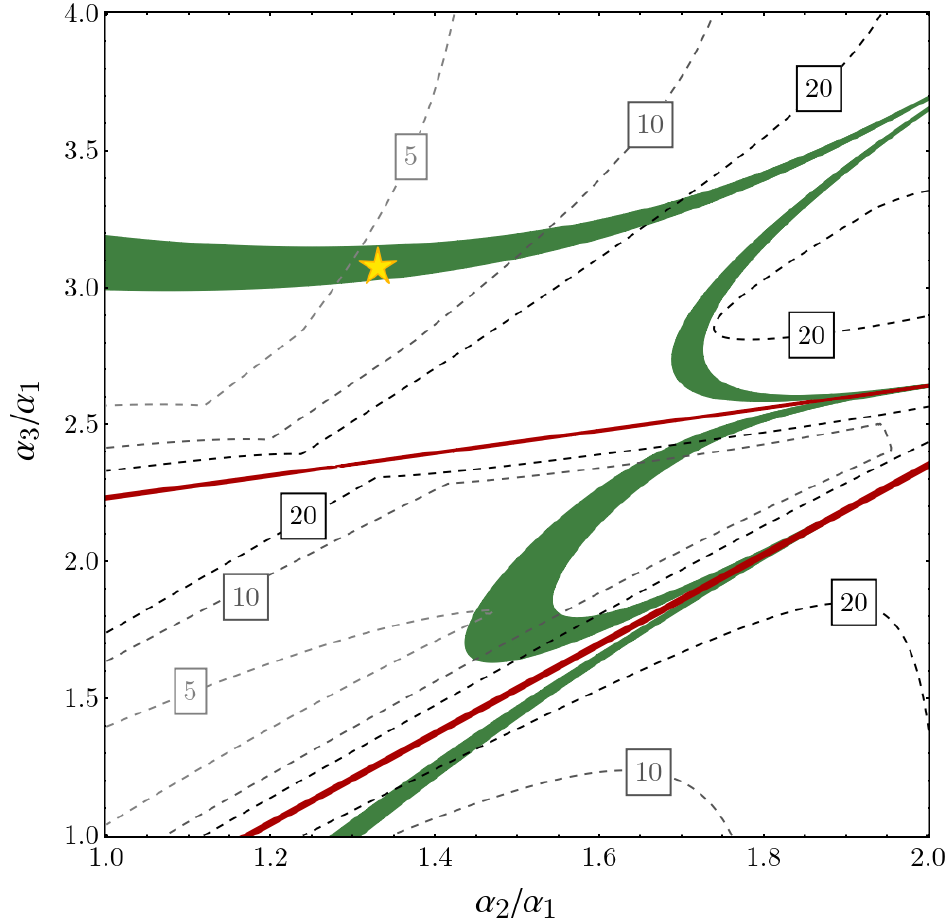}
  \caption{Values of the charged-lepton Yukawa couplings of the \(A_5'\) model with large \(\im\tau\) which, when eq.~\eqref{eq:A5_3x3'_CL_ratios} is applicable, allow to reproduce charged-lepton mass ratios at \(1\sigma\) (green). The red regions are not accessible due to an upper limit on \(|\epsilon|\) within the fundamental domain. Contours refer to a Barbieri-Giudice measure of fine-tuning (see text). The yellow star shows the location of the best-fit point for this model.
  }
  \label{fig:A5}
\end{figure}

We have performed a systematic scan restricting ourselves to models requiring modular forms of weight not higher than \(k=6\), and involving at most 8 effective parameters (including \(\re \tau\) and \(\im \tau\)). Models producing a massless electron are rejected.
For neutrino masses generated via a type I seesaw, we have considered gauge-singlet superfields \(N^c\) furnishing a complete irrep of dimension 2 or 3.
Out of the 36 models thus identified, we have selected the only one which i) is viable in the regime of interest, ii) produces a charged-lepton spectrum which is not fine-tuned,%
\footnote{Note that fitting simultaneously mass ratio and mixing angle data may drive the model parameters to tuned values, even if no tuning is expected at the level of charged-lepton masses.}
and iii) is consistent with the experimental bound on the Dirac CPV phase \(\delta\).
For this model, \(k_L =3\), \(k_E=1\) and \(N^c \sim (\mathbf{\hat{2}}', 2)\). The corresponding superpotential reads:
\begin{equation}
  \label{eq:A5_3x3'_W}
  \begin{split}
    W &= \left[
      \alpha_1 \left( Y^{(5,4)}_{\mathbf{4}} E^c L \right)_{\mathbf{1}} +
      \alpha_2 \left( Y^{(5,4)}_{\mathbf{5}, 1} E^c L \right)_{\mathbf{1}} +
      \alpha_3 \left( Y^{(5,4)}_{\mathbf{5}, 2} E^c L \right)_{\mathbf{1}}
    \right] H_d \\
    &+ \left[
      g_1 \left( Y^{(5,5)}_{\mathbf{\hat{6}}, 1} N^c L \right)_{\mathbf{1}} +
      g_2 \left( Y^{(5,5)}_{\mathbf{\hat{6}}, 2} N^c L \right)_{\mathbf{1}} +
      g_3 \left( Y^{(5,5)}_{\mathbf{\hat{6}}, 3} N^c L \right)_{\mathbf{1}}
    \right] H_u \\
    &+ \Lambda \left( Y^{(5,4)}_{\mathbf{3}'} (N^c)^2 \right)_{\mathbf{1}} \,.
  \end{split}
\end{equation}
The modular forms entering \(W\) are obtained from the lowest weight (\(k=1\)) sextet~\cite{Yao:2020zml},
\begin{equation}
    Y^{(5,1)}_{\mathbf{\hat{6}}} =
    \left(2\, \varepsilon^5+\theta^5,\,
    2 \,\theta^5-\varepsilon^5,\,
    5 \,\varepsilon\,  \theta^4,\,
    5 \sqrt{2}\, \varepsilon^2\, \theta^3,\,
    -5 \sqrt{2}\, \varepsilon^3\, \theta^2,\,
    5 \,\varepsilon^4\, \theta \right)^T\,,
\end{equation}
where \(\theta\) and  \(\varepsilon\) are functions of the modulus, \(\theta(\tau) = 1 + 3q^5/5 + 2q^{10}/25 -28q^{15}/125+\ldots\) and 
\(\varepsilon(\tau) = q\left(1- 2q^5/5 + 12q^{10}/25 +37q^{15}/125+\ldots\right)\), with \(q = \exp \left( 2 \pi i \tau / 5 \right)\). In our regime of interest, \(|q|\) is small and the sextet modular multiplet approximately reads \(Y^{(5,1)}_{\mathbf{\hat{6}}} \simeq (1,2,5q,0,0,0)\), which motivates an alternative definition of expansion parameter, \(|\epsilon|\) with \(\epsilon \equiv 5 \varepsilon / \theta \simeq 5 q\), used only in the context of this section. The charged-lepton mass matrix is then approximated by
\begin{equation}
  \label{eq:A5_3x3'_Me}
\small
\arraycolsep=0.3\arraycolsep
  M_e^\dagger \simeq   
\frac{4 \sqrt{2}}{\sqrt{3}} v_d \alpha_1 \theta^4
  \begin{pmatrix}
    \frac{9}{2} \left( \tilde{\alpha}_2 - \frac{10}{9} \tilde{\alpha}_3 \right) & \frac{\sqrt{2}}{5} \left( -\frac{14}{25} + \frac{53}{50} \tilde{\alpha}_2 - \tilde{\alpha}_3 \right) \epsilon^4 & \frac{1}{\sqrt{2}} \left( 4 + \tilde{\alpha}_2 \right) \epsilon \\[2mm]
    \frac{2 \sqrt{2}}{5} \left( -\frac{7}{5} + \tilde{\alpha}_2 - \tilde{\alpha}_3 \right) \epsilon^3 & - \left( \frac{2}{5} + \tilde{\alpha}_2 - 2 \tilde{\alpha}_3 \right) \epsilon^2 & - \frac{2}{5} \left( \frac{7}{25} + \frac{53}{50} \tilde{\alpha}_2 - \tilde{\alpha}_3 \right) \epsilon^4 \\[2mm]
    \frac{1}{\sqrt{2}} \left( - \frac{4}{5} + \tilde{\alpha}_2 - 2 \tilde{\alpha}_3 \right) \epsilon^2 & \left( 2 - \tilde{\alpha}_2 \right) \epsilon & - \frac{4}{5} \left( \frac{7}{10} + \tilde{\alpha}_2 -\tilde{\alpha}_3 \right) \epsilon^3
  \end{pmatrix} \,,
\end{equation}
with \(\tilde{\alpha}_2 \equiv \frac{2}{\sqrt{5}} \alpha_2 / \alpha_1\), \(\tilde{\alpha}_3 \equiv \frac{2 \sqrt{6}}{5 \sqrt{5}} \alpha_3 / \alpha_1\), matching the pattern in eq.~\eqref{eq:example}.
It follows that the charged-lepton mass ratios are given by
\begin{equation}
  \label{eq:A5_3x3'_CL_ratios}
  \begin{split}
    \frac{m_e}{m_{\mu}} &\simeq \frac{16}{125} \left|
      \frac{
        \left( 4 + \tilde{\alpha}_2 - 5 \tilde{\alpha}_3 \right)
        \left( 10 + 7 \tilde{\alpha}_2 - 5 \tilde{\alpha}_3 \right)
        \left( 2 - 4 \tilde{\alpha}_2 + 5 \tilde{\alpha}_3 \right)
      }{
        \left( 2 - \tilde{\alpha}_2 \right)^2
        \left( 9 \tilde{\alpha}_2 - 10 \tilde{\alpha}_3 \right)
      } \right| |\epsilon|^3 \,, \\[2mm]
    \frac{m_{\mu}}{m_{\tau}} &\simeq 2 \left|
      \frac{2 - \tilde{\alpha}_2}{9 \tilde{\alpha}_2 - 10 \tilde{\alpha}_3}
    \right| |\epsilon| \,,
  \end{split}
\end{equation}
at leading order in \(|\epsilon|\).
These expressions alone isolate viable (\(\epsilon\)-independent) regions in the plane of \(\tilde\alpha_2\) and \(\tilde\alpha_3\). Taking the \(1\sigma\) ranges for charged-lepton mass ratios from Table~\ref{tab:globalFit}, we plot these regions in Figure~\ref{fig:A5}. The superimposed contours refer to the Barbieri-Giudice measure of fine-tuning~\cite{Barbieri:1987fn} in the charged-lepton sector, \(\max(\text{BG})\), corresponding to the largest of four quantities \( |\partial \ln (\text{mass ratio})/ \partial \ln \tilde\alpha_{2,3}|\).
An observable \(O\) is typically considered fine-tuned with respect to some parameter \(p\) if \(\text{BG} \equiv |\partial \ln O / \partial \ln p| \gtrsim 10\)~\cite{Barbieri:1987fn}.

%%%%%%%%%%%%%%%%
%%%%%%%%%%%%%%%%

Expansions similar to~\eqref{eq:A5_3x3'_Me} for the neutrino Yukawa and mass matrices are not useful since \(|\epsilon|\) is not the only small parameter in the neutrino sector. In particular, some entries of \(M_{\nu}\) are proportional to \(1 + (5/12) g_2 / g_1 -\sqrt{3}\, g_3 / g_1\) which is forced by the fit to be \(\mathcal{O}(10^{-2})\) in the viable region.
Fine-tuning in the neutrino sector is expected (see section~\ref{sec:natural}) and is related to the fact that the residual symmetry constrains not only the charged-lepton masses, but also the lepton mixing pattern. By sending \(\im \tau \to \infty\) while keeping the couplings fixed to their best-fit values, one can check that the (13), (23), (31) and (32) entries of the PMNS matrix become zero. In particular, \(\sin^2 \theta_{23} \to 0\), and therefore the symmetric limit does not provide an approximate description of leptonic mixing.

The result of the fit of this \(A_5'\) model is summarised in the first column of Table~\ref{tab:pheno}, which includes best-fit values and the corresponding \(3\sigma\) ranges. The viable region in the \(\tau\) plane is shown on the left side of Figure~\ref{fig:tau} and corresponds to a neutrino spectrum with inverted ordering (IO).

%%%%%%%%%%%%%%%%%%%%%%%
\subsubsection{\texorpdfstring{\(S_4'\)}{S4'} models with \texorpdfstring{\(L \sim \mathbf{\hat{2}} \oplus  \mathbf{\hat{1}}\), \(E^c \sim \mathbf{\hat{3}}'\)}{L= 2\textasciicircum+1\textasciicircum, Ec = 3\textasciicircum'}}
\label{sec:models_S4_inf}
%%%%%%%%%%%%%%%%%%%%%%%
%

In the second most `structured' case, one of the fields \(L\), \(E^c\) is an irreducible triplet, while the other decomposes into a doublet and a singlet of the finite modular group.
This possibility is realised at level \(N = 4\) in the vicinity of \(\tau = i \infty\), see Table~\ref{tab:good_patterns}.
For definiteness, we take \(L = L_{12} \oplus L_3\) with \(L_{12} \sim (\mathbf{\hat{2}}, k_L)\), \(L_3 \sim (\mathbf{\hat{1}}, k_L)\), and \(E^c \sim (\mathbf{\hat{3}'}, k_E)\).
The charged-lepton mass pattern in this regime is predicted to be \(\left( m_{\tau}, m_{\mu}, m_e \right) \sim (1, \epsilon, \epsilon^3)\).

\begin{figure}[t!]
  \centering
  \includegraphics[width=0.65\textwidth]{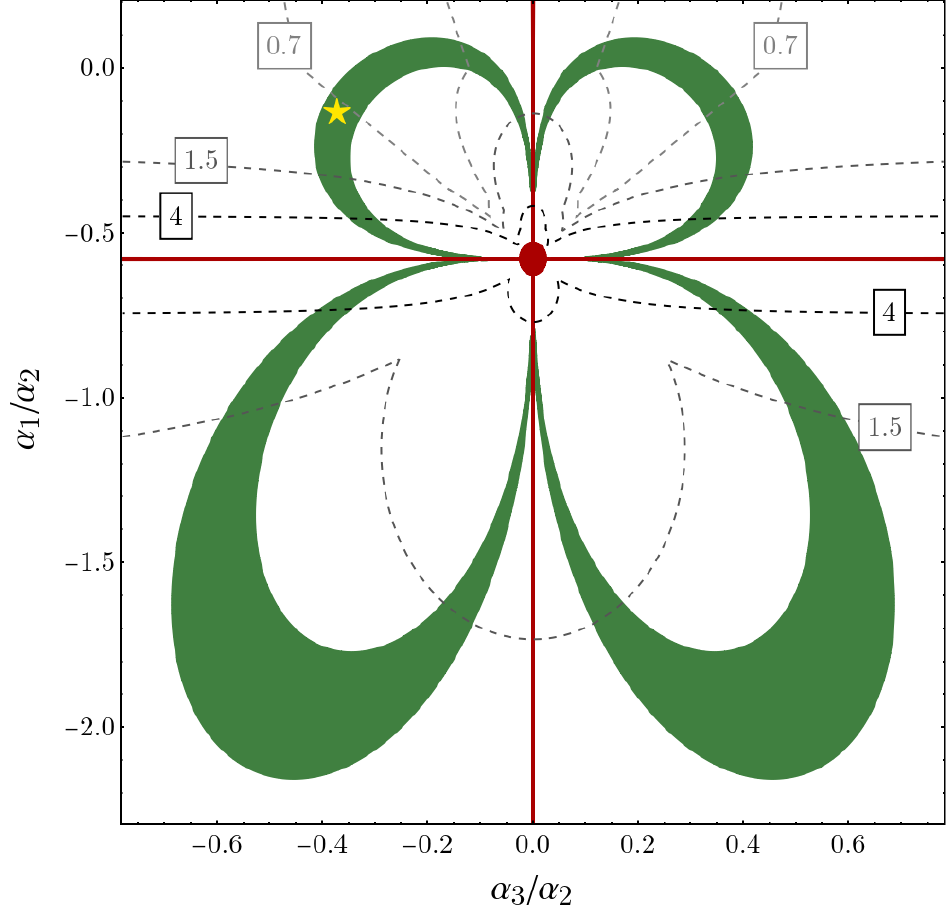}
  \caption{
  Values of the charged-lepton Yukawa couplings of the \(S_4'\) model with large \(\im\tau\) which, when eq.~\eqref{eq:S4_inf_CL_ratios} is applicable, allow to reproduce charged-lepton mass ratios at \(1\sigma\) (green). The red regions are not accessible due to an upper limit on \(|\epsilon|\) within the fundamental domain. Contours refer to the Barbieri-Giudice measure of fine-tuning (see text). The yellow star shows the location of the best-fit point for this model.}
  \label{fig:S4}
\end{figure}

We have performed a systematic scan restricting ourselves to models involving at most 8 effective parameters (including \(\re \tau\) and \(\im \tau\); no limit on modular form weights). Once again, models predicting a massless electron are rejected, while the \(N^c\) (when present) furnish a complete irrep of dimension 2 or 3.
Out of the 60 models thus identified, we have selected the only one which i) is viable in the regime of interest and ii) produces a charged-lepton spectrum which is not fine-tuned. This model turns out to be consistent with the experimental bound on the Dirac CPV phase. It corresponds to \(k_L = k_E=2\) and \(N^c \sim (\mathbf{3}, 1)\) and the superpotential reads:
\begin{equation}
  \label{eq:S4_inf_W}
  \begin{aligned}
    W &= \left[
      \alpha_1 \left( Y^{(4,4)}_{\mathbf{3}} E^c L_{12} \right)_{\mathbf{1}} +
      \alpha_2 \left( Y^{(4,4)}_{\mathbf{3'}} E^c L_{12} \right)_{\mathbf{1}} +
      \alpha_3 \left( Y^{(4,4)}_{\mathbf{3}} E^c L_{3} \right)_{\mathbf{1}}
    \right] H_d \\
    &+ \left[
      g_1 \left( Y^{(4,3)}_{\mathbf{\hat{3}}} N^c L_{12} \right)_{\mathbf{1}} +
      g_2 \left( Y^{(4,3)}_{\mathbf{\hat{3}'}} N^c L_{12} \right)_{\mathbf{1}} +
      g_3 \left( Y^{(4,3)}_{\mathbf{\hat{3}'}} N^c L_3 \right)_{\mathbf{1}}
    \right] H_u \\
    &+ \Lambda \left( Y^{(4,2)}_{\mathbf{2}} (N^c)^2 \right)_{\mathbf{1}} \,.
  \end{aligned}
\end{equation}
The modular forms entering \(W\) are obtained from the lowest weight (\(k=1\)) triplet~\cite{Novichkov:2020eep},
\begin{equation}
    Y^{(4,1)}_{\mathbf{\hat{3}}} =
    \left( \sqrt{2} \varepsilon \theta,\,
    \varepsilon^2,\,
    -\theta^2 \right)^T\,,
\end{equation}
where \(\theta(\tau) = 1 + 2q^4 + 2q^{16}+\ldots\) and \(\varepsilon(\tau) = 2 q  + 2q^9 + 2q^{25}+\ldots\), with \(q = \exp \left( i \pi \tau / 2 \right)\). Using in the context of this section the expansion parameter obtained from the ratio of these functions, \(|\epsilon|\) with \(\epsilon \equiv \varepsilon / \theta \simeq 2 q\), the charged-lepton mass matrix is approximately given by
\begin{equation}
  \label{eq:S4_inf_Me}
  M_e^\dagger \simeq
  \frac{\sqrt{3}}{2}  v_d \alpha_1 \theta^8
  \begin{pmatrix}
 \epsilon^2 & \frac{\left(\tilde{\alpha}_2+\sqrt{3}\right) }{2 \sqrt{6}} \epsilon & \frac{\left(7 \tilde{\alpha}_2-\sqrt{3}\right) }{2 \sqrt{6}} \epsilon ^3\\[2mm]
 -\frac{\tilde{\alpha}_2}{6} & \frac{\left(7 \sqrt{3} \tilde{\alpha}_2+9\right) }{6 \sqrt{6}} \epsilon ^3& \frac{\left(\sqrt{3} \tilde{\alpha}_2-9\right)  }{6 \sqrt{6}} \epsilon\\[2mm]
 \tilde{\alpha}_3 \epsilon ^2 & -\frac{\tilde{\alpha}_3}{\sqrt{2}} \epsilon  & \frac{\tilde{\alpha}_3 }{\sqrt{2}}\epsilon ^3
 \end{pmatrix}\,,
\end{equation}
with \(\tilde{\alpha}_2 \equiv \alpha_2 / \alpha_1\) and \(\tilde{\alpha}_3 \equiv \alpha_3 / \alpha_1\). It matches the expected power structure in \(|q|\), as one can check.%
\footnote{Aside from consulting the third column of Table~\ref{tab:S4p_residual_reps}, one must keep in mind the ordering of the \(\rho_i^{(c)}\), which depends on the representation basis for the group generators (we are using that of~\cite{Novichkov:2020eep}).}
One can also find approximate expressions for the charged-lepton mass ratios, which read
\begin{equation}
  \label{eq:S4_inf_CL_ratios}
  \begin{aligned}
    \frac{m_e}{m_{\mu}} &\simeq 18 \sqrt{3} \frac{
      \left| \tilde{\alpha}_3 (\tilde{\alpha}_2^2 - 3) \right|
    }{
    |\tilde{\alpha}_2| \left( (\tilde{\alpha}_2 + \sqrt{3})^2 + 12 \tilde{\alpha}_3^2 \right)} |\epsilon|^2 \,, \\[2mm]
    \frac{m_{\mu}}{m_{\tau}} &\simeq \sqrt{\frac{3}{2}} \frac{\sqrt{(\tilde{\alpha}_2 + \sqrt{3})^2 + 12 \tilde{\alpha}_3^2}}{|\tilde{\alpha}_2|} |\epsilon| \,.
  \end{aligned}
\end{equation}
The expressions~\eqref{eq:S4_inf_CL_ratios} isolate viable (\(\epsilon\)-independent) regions in the plane of coupling constants, say~\(\tilde\alpha_2^{-1} = \alpha_1 / \alpha_2\) and \(\tilde\alpha_3/\tilde\alpha_2 = \alpha_3 / \alpha_2\). We plot these regions in Figure~\ref{fig:S4}, including contours quantifying the degree of fine-tuning involved in the relation between charged-lepton mass ratios and superpotential parameters (as described in the previous section). Note that the model best-fit point in particular corresponds to a small value of \(\max(\text{BG}) \simeq 0.74\).
%
%%%%%%%%%%%%%%%%
%%%%%%%%%%%%%%%%

Regarding the neutrino sector, one can check that close to the symmetric limit the neutrino masses follow the pattern \((a |\epsilon|^6, b_0 - b_2 |\epsilon|^2, b_0 + b_2 |\epsilon|^2)\), which naturally leads to IO. However, we were unable to find viable regions with IO and, instead, the model predicts a neutrino spectrum with normal ordering (NO). Within the viable region, the approximate pattern \((a |\epsilon|^6, b_0 - b_2 |\epsilon|^2, b_0 + b_2 |\epsilon|^2)\) is not accurate since \(|\epsilon|\) is not the only small parameter in the neutrino sector. In particular, some of the entries of \(M_{\nu}\) are proportional to \((1 + \sqrt{6} g_2 / g_3)\) which is forced to be \(\mathcal{O}(10^{-3})\) in our fit.
As in the previous section, the fine-tuning in the neutrino sector is explained by the necessity to introduce large corrections to the symmetric-limit PMNS matrix, which has a zero either in the (31) or (33) entry in the case of NO or IO, respectively.

The result of the fit of this \(S_4'\) model is summarised in the second column of Table~\ref{tab:pheno}. The viable region in the \(\tau\) plane is located near the point \(\tau = 2.7 i\) and is shown on the left side of Figure~\ref{fig:tau}.

%%%%%%%%%%%%%%%%%%%%%%%
\section{Large mixing angles without fine-tuning}
\label{sec:large_mixing}
%%%%%%%%%%%%%%%%%%%%%%%
%

%%%%%%%%%%%%%%%%%%%%%%%
\subsection{Viable PMNS matrix in the symmetric limit}
\label{sec:natural}
%%%%%%%%%%%%%%%%%%%%%%%
%

We have seen in the previous sections that a slightly broken residual modular symmetry allows to accommodate hierarchical charged-lepton masses without fine-tuning of the corresponding couplings.
However, the resulting models are still subject to fine-tuning in the neutrino sector, since residual symmetries typically constrain not only the charged-lepton masses, but also the form of the PMNS matrix by forcing some of its entries to be zeros.
This raises the question of whether it is possible to have a PMNS matrix which is close to the observed one even in the symmetric limit, i.e.~such that either none of its entries vanish, or only the (13) entry vanishes as \(\epsilon \to 0\).

This possibility has been investigated in Ref.~\cite{Reyimuaji:2018xvs} for arbitrary flavour symmetry groups.
In particular, this analysis directly applies to the case of the flavour symmetry being a residual modular symmetry.
One of the main conclusions of Ref.~\cite{Reyimuaji:2018xvs} is that only a limited number of flavour symmetry representation choices for \(L\) and \(E^c\) give rise to a PMNS matrix which is viable in the symmetric limit (as defined above).
Most notably, there are only two such cases consistent with hierarchical charged-lepton masses:
\begin{enumerate}
\item \(L \leadsto 1 \oplus 1 \oplus 1\), \(E^c \leadsto 1 \oplus r\), where \(1\) is some real singlet of the flavour symmetry, and \(r\) is some (possibly reducible) representation such that \(r \not\supset 1\);
\item \(L \leadsto \mathbf{1} \oplus \mathbf{1} \oplus \mathbf{1}^*\), \(E^c \leadsto \mathbf{1}^* \oplus r\), where \(\mathbf{1}\) is some complex singlet of the flavour symmetry, \(\mathbf{1}^*\) is its conjugate, and \(r\) is some (possibly reducible) representation such that \(r \not\supset \mathbf{1}, \mathbf{1}^*\).
\end{enumerate}

The above original result makes use of the assumption that one charged-lepton mass and at least one neutrino mass does not vanish in the symmetric limit. However, one can also deduce from the analysis performed in~\cite{Reyimuaji:2018xvs} that the PMNS matrix is generically unconstrained in the symmetric limit when the opposite is true.
Therefore, we extend the list of viable cases with the following two:
\begin{enumerate}
\setcounter{enumi}{2}
\item all charged-lepton masses vanish in the symmetric limit, i.e.~the corresponding hierarchical pattern involves only positive powers of \(\epsilon\), e.g. \((\epsilon, \epsilon^2, \epsilon^3)\);
\item all light neutrino masses vanish in the symmetric limit, i.e.~\(L\) decomposes into three (possibly identical) complex singlets none of which are conjugated to each other.
\end{enumerate}

It follows that a modular-symmetric model of lepton flavour with hierarchical charged-lepton masses may be free of fine-tuning if it satisfies any of the properties 1-4.
Applying this filter to the promising hierarchical cases of Table~\ref{tab:good_patterns}, one is left with the representation pairs listed here, in Table~\ref{tab:the_greenest_table}.
In this summary table, we have once again disregarded reducible representations made up of three copies of the same singlet.
%
%%%%%%%%%%%%%%%%%%%%%%%
\begin{table}[ht!]
\small
  \centering
  \renewcommand{\arraystretch}{1.5}
  \begin{tabular}{cccllc}
\toprule
\belowrulesepcolor{light-gray}
\rowcolor{light-gray}
\(\,N\,\) & \(\,\Gamma'_N\,\) & \(\,\)Pattern\(\,\) & Sym.~point & Viable \(\mathbf{r}_{E^c} \otimes \mathbf{r}_{L}\quad\)  & \,\,Case\,\,\\
\aboverulesepcolor{light-gray}
\midrule
%%%%%%%%%%%%%%
  2 
& \(S_3\) 
& \((1,\epsilon,\epsilon^2)\)                
& \(\tau \simeq \omega\)  
& \([\mathbf{2}\oplus\mathbf{1}^{(\prime)}] \otimes [\mathbf{1}\oplus\mathbf{1}^{(\prime)}\oplus\mathbf{1}']\) 
& 1 or 4
\\[2mm]
%%%%%%%%%%%%%%
\rowcolor{lighter-gray}
& &
& \(\tau \simeq \omega\)
& \([\mathbf{1}_a\oplus\mathbf{1}_a\oplus\mathbf{1}_a'] \otimes [\mathbf{1}_b\oplus\mathbf{1}_b\oplus\mathbf{1}_b'']\)
& 2
\\[1mm]
\rowcolor{lighter-gray}
& & &
& \([\mathbf{1}\oplus\mathbf{1}\oplus\mathbf{1}'] \otimes [\mathbf{1}''\oplus\mathbf{1}''\oplus\mathbf{1}']\), %or
&
\\[-1mm]
\rowcolor{lighter-gray}
\multirow{-2.7}{*}{3}
& \multirow{-2.7}{*}{\(A_4'\) }
& \multirow{-2.7}{*}{\((1,\epsilon,\epsilon^2)\)}
& \multirow{-1.7}{*}{\(\tau \simeq i \infty\)}
& \([\mathbf{1}\oplus\mathbf{1}\oplus\mathbf{1}''] \otimes 
[\mathbf{1}'\oplus\mathbf{1}'\oplus\mathbf{1}'']\)
& \multirow{-1.6}{*}{2}
\\[2mm]
%%%%%%%%%%%%%%
4
& \(S_4'\)
& \((1,\epsilon,\epsilon^2)\) 
& \(\tau \simeq \omega\)
& 
\([
\mathbf{3}_a \text{, or }
\mathbf{2}\oplus\mathbf{1}^{(\prime)} \text{, or }
\mathbf{\hat{2}}\oplus\mathbf{\hat{1}}^{(\prime)}
] \otimes [\mathbf{1}_b\oplus\mathbf{1}_b\oplus\mathbf{1}_b']\)
& 1 or 4
\\[2mm]
%%%%%%%%%%%%%%
\rowcolor{lighter-gray}
5
& \(A_5'\)
& \(-\) 
& \(-\) 
& \(-\)   
& \(-\)   
\\
\aboverulesepcolor{lighter-gray}
\bottomrule
  \end{tabular}
  \caption{
  Hierarchical charged-lepton mass patterns which may be realised in the vicinity of symmetric points without fine-tuned mixing (PMNS close to the observed one in the symmetric limit). The property which is satisfied (from 1-4, see text) is given in the last column and may depend on the weights \(k\) and \(k^c\). The case \(N=3\) with \(\tau \simeq \omega\) is the only one in the table for which \(\mathbf{r}_{E^c}\leftrightarrow \mathbf{r}_L\) may be required, and for which not all \(k^{(c)}\) choices are viable. For other notation, see the caption of Table~\ref{tab:good_patterns}.}
  \label{tab:the_greenest_table}
\end{table}
%%%%%%%%%%%%%%%%%%%%%%%
%

We now proceed by constructing such a model in the following section (clearly, the models described in section~\ref{sec:models_inf} do not satisfy any of the properties 1-4).
As a final remark, we note that the argument of Ref.~\cite{Reyimuaji:2018xvs} is only valid in the case when the flavour symmetry analysis can be applied directly to the light neutrino mass matrix.
In our setup, this corresponds to the situation when light neutrino masses arise either directly from a modular-invariant Weinberg operator, or via a type-I seesaw UV completion such that none of the gauge-singlet neutrinos \(N^c\) becomes massless in the symmetric limit (so that they can be integrated out). This is the case in the two models considered so far and in the model described in the following section.

%%%%%%%%%%%%%%%%%%%%%%%
\subsection{\texorpdfstring{\(S_4'\)}{S4'} models with \texorpdfstring{\(\tau \simeq \omega\)}{tau close to omega}}
\label{sec:models_S4_omega}
%%%%%%%%%%%%%%%%%%%%%%%
%

We finally turn to the most `structured' cases within the surviving lepton flavour models of Table~\ref{tab:the_greenest_table}. These arise at level~\(N = 4\) in the vicinity of~\(\tau = \omega\) and correspond to \(E^c\) and \(L\) being a triplet and the direct sum of three singlets of the finite modular group \(S_4'\), respectively. 
The expected charged-lepton mass pattern is \((m_{\tau}, m_{\mu}, m_e) \sim (1, \epsilon, \epsilon^2)\).

We have performed a systematic scan restricting ourselves to promising models involving the minimal number of effective parameters (9, including \(\re \tau\) and \(\im \tau\)). Once again, models predicting a massless electron are rejected, while the \(N^c\) furnish a complete irrep of dimension 2 or 3 (\(N^c\) are present since Weinberg models require more parameters).
Out of 48 models, we have identified the only one which i) is viable in the regime of interest, ii) is not fine-tuned in this regime, and iii) is consistent with the \(2\sigma\) range for the Dirac CPV phase, predicting \(\delta \simeq \pi\) while other models lead to \(\delta \simeq 0\).
For this model, \(L = L_1 \oplus L_2 \oplus L_3\) with \(L_1, L_2 \sim (\mathbf{\hat{1}}, 2)\), \(L_3 \sim (\mathbf{\hat{1}'}, 2)\), and \(E^c \sim (\mathbf{\hat{3}}, 4)\) and \(N^c \sim (\mathbf{3}', 1)\). The corresponding superpotential reads:
\begin{equation}
  \label{eq:S4_omega_W}
  \begin{aligned}
    W &= \left[
      \alpha_1 \left( Y^{(4,6)}_{\mathbf{3'}, 1} E^c L_1 \right)_{\mathbf{1}} +
      \alpha_2 \left( Y^{(4,6)}_{\mathbf{3'}, 2} E^c L_1 \right)_{\mathbf{1}} \right. \\
      &+ \left. \alpha_3 \left( Y^{(4,6)}_{\mathbf{3'}, 1} E^c L_2 \right)_{\mathbf{1}} +
      \alpha_4 \left( Y^{(4,6)}_{\mathbf{3'}, 2} E^c L_2 \right)_{\mathbf{1}} +
      \alpha_5 \left( Y^{(4,6)}_{\mathbf{3}} E^c L_3 \right)_{\mathbf{1}}
    \right] H_d \\
    &+ \left[
      g_1 \left( Y^{(4,3)}_{\mathbf{\hat{3}}} N^c L_1 \right)_{\mathbf{1}} +
      g_2 \left( Y^{(4,3)}_{\mathbf{\hat{3}}} N^c L_2 \right)_{\mathbf{1}} +
      g_3 \left( Y^{(4,3)}_{\mathbf{\hat{3}'}} N^c L_3 \right)_{\mathbf{1}}
    \right] H_u \\
    &+ \Lambda \left( Y^{(4,2)}_{\mathbf{2}} (N^c)^2 \right)_{\mathbf{1}} \,.
  \end{aligned}
\end{equation}
Since \(L_1\) and \(L_2\) are indistinguishable, one of the constants \(\alpha_i\), with \(i = 1, \dots, 4\), is effectively not an independent parameter and can be set to zero by a suitable rotation without loss of generality. We choose to set \(\alpha_2 = 0\).

At leading order in the small parameter \(|\epsilon|\), with \(\epsilon \equiv 1 - \frac{1 + \sqrt{3}}{1 - i} \frac{\varepsilon}{\theta}\) and \(|\epsilon| \simeq 2.8 \left| \frac{\tau - \omega}{\tau - \omega^2} \right|\) in the context of this section,%
\footnote{
  The definition of \(\epsilon\) is motivated by the fact that \(\varepsilon/\theta = (1-i)/(1+\sqrt{3})\) at \(\tau = \omega\), see eq.~(3.5) in Ref.~\cite{Novichkov:2020eep} (note the wrong sign in front of \(i\) in the published version of~\cite{Novichkov:2020eep}).
}
the charged-lepton mass matrix reads
\begin{equation}
  \label{eq:S4_cusp_Me}
  M_e^\dagger \simeq
  -\frac{3 (\sqrt{3}-1)^6}{\sqrt{13}}
  v_d \alpha_1 \theta^{12}
  \begin{pmatrix}
 1 & \tilde\alpha_3+\frac{\sqrt{13}}{2}\tilde\alpha_4 & \frac{i\sqrt{39}}{2} \tilde\alpha_5 \\[2mm]
 \sqrt{3} \,\epsilon  
 & \sqrt{3} \left(\tilde\alpha_3 -\frac{\sqrt{13}}{2}  \tilde\alpha_4\right) \epsilon  
 & \frac{i\sqrt{13}}{2}  \tilde\alpha_5 \,\epsilon  \\[2mm]
 \frac{5}{2}\, \epsilon^2 & \frac{1}{4} \left(10 \tilde\alpha _3+\sqrt{13} \tilde\alpha_4\right) \epsilon^2 & -\frac{5i\sqrt{13}}{4\sqrt{3}} \tilde\alpha_5 \, \epsilon^2
 \end{pmatrix}\,,
\end{equation}
while the charged-lepton mass ratios are given by
\begin{equation}
  \label{eq:S4_omega_CL_ratios}
  \begin{split}
    \frac{m_e}{m_{\mu}} &\simeq 2 \frac{
      | \tilde{\alpha}_4 \tilde{\alpha}_5 | \sqrt{4 + \left( 2 \tilde{\alpha}_3 + \sqrt{13} \tilde{\alpha}_4 \right)^2 + 39 \tilde{\alpha}_5^2}
    }{
    3 \tilde{\alpha}_4^2 + \left[ 1 + \left( \tilde{\alpha}_3 - \sqrt{13} \tilde{\alpha}_4 \right)^2 \right] \tilde{\alpha}_5^2} \,|\epsilon| \,, \\[2mm]
  \frac{m_{\mu}}{m_{\tau}} &\simeq 4 \sqrt{13} \frac{
    \sqrt{3 \tilde{\alpha}_4^2 + \left[ 1 + \left( \tilde{\alpha}_3 - \sqrt{13} \tilde{\alpha}_4 \right)^2 \right] \tilde{\alpha}_5^2}
  }{4 + \left( 2 \tilde{\alpha}_3 + \sqrt{13} \tilde{\alpha}_4 \right)^2 + 39 \tilde{\alpha}_5^2} \,|\epsilon| \,,
  \end{split}
\end{equation}
with \(\tilde{\alpha}_i \equiv \alpha_i / \alpha_1\), \(i = 3, 4, 5\). With respect to charged-lepton mass ratios, the model best-fit point is found to correspond to \(\max(\text{BG}) \simeq 0.85\).

Up to an overall normalisation \(\mathcal{K}\), the light neutrino mass matrix is instead given by:
\begin{equation}
  \label{eq:S4_cusp_Mnu}
  M_{\nu} \,\simeq\, \mathcal{K}\,\epsilon
  \begin{pmatrix}
    0 & 0 & \tilde{g}_3 \\
    0 & 0 & \tilde{g}_2 \tilde{g}_3 \\
    \tilde{g}_3 & \tilde{g}_2 \tilde{g}_3 & 2i \sqrt{\frac{2}{3}} \tilde{g}_3^2
  \end{pmatrix}
\end{equation}
at leading order in \(|\epsilon|\), where \(\tilde{g_i} \equiv g_i / g_1\), \(i = 2, 3\).
Note that the smallness of \(|\epsilon|\) does not constrain the \(M_\nu\) contribution to the mixing matrix, which depends only on the couplings \(g_i\), and large mixing angles are allowed.

\begin{figure}[ht!]
  \centering
  \includegraphics[width=0.62\textwidth]{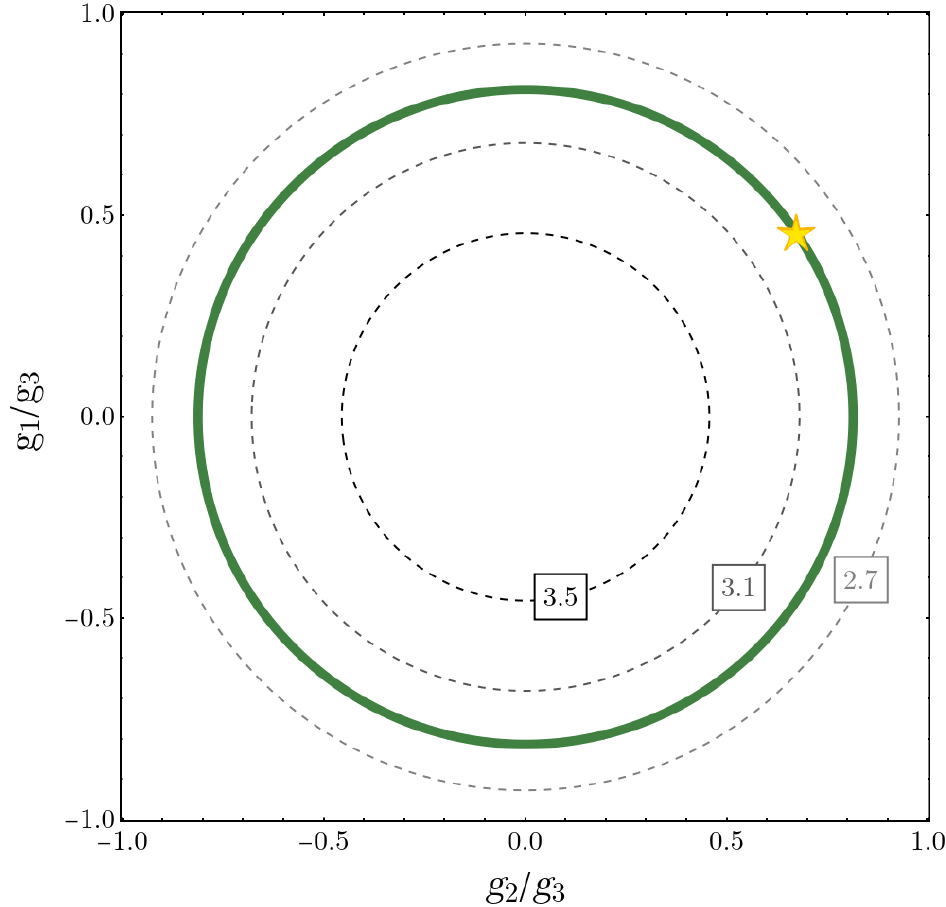}
  \caption{
    Values of the neutrino Yukawa couplings of the \(S_4'\) model with \(\tau\simeq \omega\) which, when eq.~\eqref{eq:S4_cusp_r} is applicable, allow to reproduce the ratio \(r\) at \(1\sigma\) (green). Contours refer to a Barbieri-Giudice measure of fine-tuning (see text). The yellow star shows the location of the best-fit point for this model.
  }
  \label{fig:S4cusp}
\end{figure}

From the form of \(M_\nu\) it is clear that, in the limit of unbroken SUSY, there is a massless neutrino, even though \(N^c\) is a triplet. This follows from the modular-symmetric superpotential, which implies the proportionality of the first two columns of \(\mathcal{Y}\), reducing its rank and therefore the rank of \(M_\nu\).
The neutrino masses thus read
\begin{equation}
  \label{eq:S4_cusp_nu_masses}
  m_1 = 0 \,, \quad
  m_{2,3} \simeq \sqrt{\frac{2}{3}}\, \mathcal{K}\, \tilde{g}_3^2 \left(\sqrt{1 + \frac{3(1 + \tilde{g}_2^2)}{2 \tilde{g}_3^2}} \mp 1 \right) |\epsilon| \,,
\end{equation}
and imply the \(\epsilon\)-independent prediction
\begin{equation}
  \label{eq:S4_cusp_r}
r = \frac{m_2^2-m_1^2}{m_3^2 -(m_1^2+m_2^2)/2}
\simeq 
\frac{6+6\tilde g_2^2+8\tilde g_3^2-4\sqrt{6\big(1+\tilde g_2^2\big)+4\tilde g_3^2}\,|\tilde g_3|}{3+3\tilde g_2^2+4\tilde g_3^2+6\sqrt{6\big(1+\tilde g_2^2\big)+4\tilde g_3^2}\,|\tilde g_3|}\,,
\end{equation}
which, by taking into account the \(1 \sigma\) range for \(r\) in Table~\ref{tab:globalFit}, isolates a viable region in the plane of coupling constants. We show this region in the plane of \(g_2/g_3\) and \(g_1/g_3\) in Figure~\ref{fig:S4cusp}. Contours refer to a Barbieri-Giudice measure of fine-tuning for the ratio of neutrino mass-squared differences, given by \( \max\{|\partial \ln r/ \partial \ln \tilde g_2|,|\partial \ln r/ \partial \ln \tilde g_3|\}\). At the model best-fit point, it has an acceptable value of \(2.9\). Additionally, the \(3\sigma\) ranges for \(\tilde g_{2,3}\) are not especially narrow. 

\begin{figure}[ht!]
  \centering
  \includegraphics[width=\textwidth]{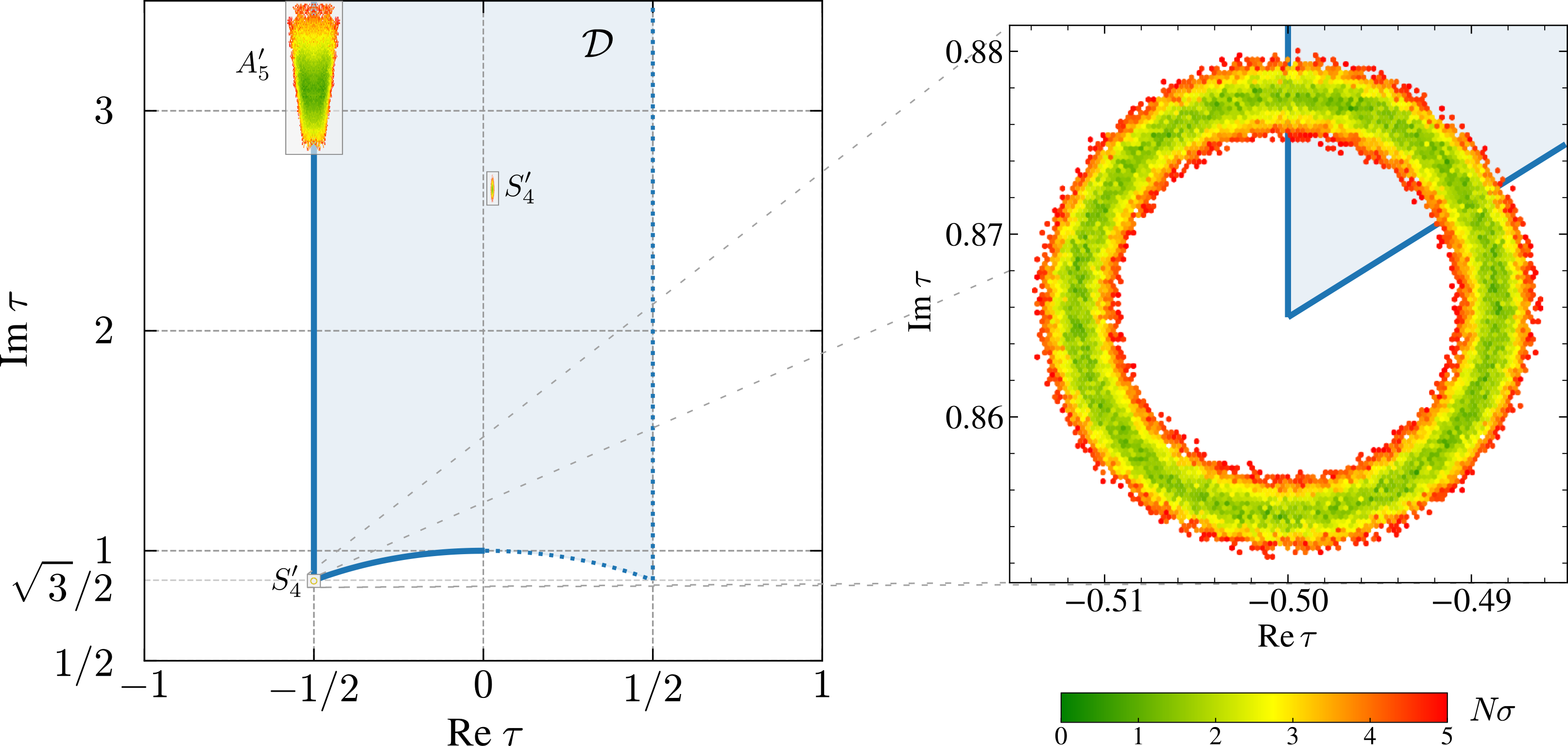}
  \caption{
  Allowed regions in the \(\tau\) plane for the models discussed in sections~\ref{sec:models_inf} and~\ref{sec:models_S4_omega} (left). The region corresponding to the model of section~\ref{sec:models_S4_omega} is magnified (right). The green, yellow and red colours correspond to different confidence levels (see legend). Points outside the fundamental domain, while redundant, are kept for illustrative purposes.}
  \label{fig:tau}
\end{figure}

The result of the fit of this \(S_4'\) model is summarised in the last column of Table~\ref{tab:pheno}. The viable region in the \(\tau\) plane corresponds to a neutrino spectrum with NO and is located very close to \(\tau_\text{sym} = \omega\), as can be seen from the magnified plot on the right side of Figure~\ref{fig:tau}.
The annular form of the region is explained by the fact that the phase of \((\tau - \omega)\) has no effect on the observables, as it enters only through \(\epsilon\) and its effects are suppressed by the smallness of \(|\epsilon|\).
Therefore, in the regime \(\tau \simeq \omega\) this model is effectively described by 8 rather than 9 parameters.

\vskip 2mm
In summary, in the vicinity of the symmetric point, i.e.~for small \(|\epsilon|\), this model can naturally lead to the observed charged-lepton mass hierarchies, see eq.~\eqref{eq:S4_omega_CL_ratios}. The neutrino mass-squared difference ratio \(r\) is, in this region, insensitive to \(\epsilon\) and depends only on the two ratios \(\tilde g_{2,3}\) of neutrino couplings, see eq.~\eqref{eq:S4_cusp_r}. Furthermore, it is not especially sensitive to these couplings.
Finally, since light neutrino masses vanish in the symmetric point, the symmetric limit allows for a generic mixing matrix (case 4 of section~\ref{sec:natural}). Therefore, the fit is not expected to be tuned in a way that compensates some `wrong PMNS' symmetric prediction. 
In fact, we have numerically verified that sending \(\tau \to \omega\) (\(\epsilon \to 0\)) has almost no effect on the values of mixing angles. This can be understood by considering, in turn, each of the contributions to the mixing matrix. 
The rotation to the mass basis in the neutrino sector, on the one hand, is independent of \(\epsilon\) in the region of interest, see eq.~\eqref{eq:S4_cusp_Mnu}, and thus has a well-defined limit as \(\epsilon \to 0\) (it is unchanged) even though light neutrinos become massless. This rotation depends only on the ratios \(\tilde g_{2,3}\) of neutrino couplings.
On the other hand, one can check that the charged-lepton rotation arising from the diagonalisation of \(M_e M_e^\dagger\), with \(M_e^\dagger\) given in eq.~\eqref{eq:S4_cusp_Me}, also has a well-defined limit as \(\epsilon \to 0\) even though two of the three charged leptons become massless. This limiting form closely matches the rotation obtained for finite, non-zero \(\epsilon\), and depends only on the ratios \(\tilde \alpha_{3,4,5}\) of charged-lepton couplings.

\vfill
\clearpage

\newgeometry{top=3cm,bottom=4.5cm}

%%%%%%%%%%%%%%%%
%%%%%%%%%%%%%%%%

%%%%%%%%%%%%%%%%
%%%%%%%%%%%%%%%%
\restoregeometry

\vfill
\clearpage

\begin{table}[ht!]
\small
  \centering
  \renewcommand{\arraystretch}{1.4}
  \begin{tabular}{lccc}
  \toprule
Model                                  &  Section~\ref{sec:models_A5_3x3'} (\(A_5'\)) &   Section~\ref{sec:models_S4_inf} (\(S_4'\)) & Section~\ref{sec:models_S4_omega} (\(S_4'\))\\
\midrule
\(\re \tau\)                             &         \(-0.47_{-0.096}^{+0.037}\) &       \(0.0235_{-0.002}^{+0.0019}\) &        \(-0.496_{-0.016}^{+0.009}\) \\
\(\im \tau\)                             &            \(3.11_{-0.19}^{+0.26}\) &            \(2.65_{-0.04}^{+0.05}\) &        \(0.877_{-0.024}^{+0.0023}\) \\
\(\alpha_2 / \alpha_1\)                  &            \(1.33_{-0.18}^{+0.20}\) &           \(-7.43_{-12.2}^{+2.76}\) &                               --- \\
\(\alpha_3 / \alpha_1\)                  &            \(3.07_{-0.15}^{+0.21}\) &            \(2.76_{-1.33}^{+5.27}\) &            \(2.45_{-0.42}^{+0.44}\) \\
\(\alpha_4 / \alpha_1\)                  &                               --- &                               --- &            \(-2.37_{-0.3}^{+0.36}\) \\
\(\alpha_5 / \alpha_1\)                  &                               --- &                               --- &            \(1.01_{-0.06}^{+0.06}\) \\
\(g_2 / g_1\)                            &     \(-0.0781_{-0.0346}^{+0.0228}\) &      \(-0.407_{-0.0003}^{+0.0002}\) &             \(1.5_{-0.14}^{+0.15}\) \\
\(g_3 / g_1\)                            &        \(0.57_{-0.0017}^{+0.0023}\) &          \(0.321_{-0.043}^{+0.02}\) &            \(2.22_{-0.15}^{+0.17}\) \\
\(v_d\,\alpha_1\), GeV                   &         \(0.404_{-0.149}^{+0.303}\) &             \(1.73_{-1.15}^{+1.8}\) &            \(4.61_{-1.33}^{+1.32}\) \\
\(v_u^2\, g_1 / \Lambda\), eV            &          \(0.778_{-0.477}^{+1.13}\) &             \(42.5_{-5.2}^{+9.88}\) &         \(0.268_{-0.063}^{+0.057}\) \\
\midrule
\(\epsilon(\tau)\)                       &      \(0.0998_{-0.0274}^{+0.0267}\) &      \(0.0313_{-0.0022}^{+0.0021}\) &      \(0.0186_{-0.0023}^{+0.0028}\) \\
CL mass pattern                          &       \((1, \epsilon, \epsilon^4)\) &       \((1, \epsilon, \epsilon^3)\) &       \((1, \epsilon, \epsilon^2)\) \\
\(\max(\text{BG})\)                      &                           \(5.579\) &                           \(0.738\) &                           \(0.848\) \\
\midrule
\(m_e / m_\mu\)                          &    \(0.00474_{-0.0005}^{+0.00062}\) &   \(0.00479_{-0.00056}^{+0.00058}\) &   \(0.00475_{-0.00052}^{+0.00061}\) \\
\(m_\mu / m_\tau\)                       &      \(0.0573_{-0.0137}^{+0.0111}\) &       \(0.0574_{-0.013}^{+0.0117}\) &      \(0.0556_{-0.0116}^{+0.0136}\) \\
\(r\)                                    &      \(0.0297_{-0.0021}^{+0.0021}\) &      \(0.0298_{-0.0023}^{+0.0019}\) &     \(0.0298_{-0.0023}^{+0.00196}\) \\
\(\delta m^2\), \(10^{-5} \text{ eV}^2\)   &             \(7.33_{-0.4}^{+0.39}\) &            \(7.38_{-0.44}^{+0.34}\) &            \(7.38_{-0.44}^{+0.35}\) \\
\(|\Delta m^2|\), \(10^{-3} \text{ eV}^2\) &            \(2.47_{-0.04}^{+0.04}\) &            \(2.48_{-0.04}^{+0.05}\) &            \(2.48_{-0.04}^{+0.05}\) \\
\(\sin^2 \theta_{12}\)                   &         \(0.306_{-0.028}^{+0.036}\) &         \(0.301_{-0.034}^{+0.044}\) &         \(0.304_{-0.036}^{+0.039}\) \\
\(\sin^2 \theta_{13}\)                   &      \(0.0222_{-0.0018}^{+0.0021}\) &      \(0.0223_{-0.0022}^{+0.0017}\) &       \(0.0221_{-0.002}^{+0.0019}\) \\
\(\sin^2 \theta_{23}\)                   &          \(0.55_{-0.097}^{+0.044}\) &         \(0.548_{-0.107}^{+0.045}\) &        \(0.539_{-0.099}^{+0.0522}\) \\
\midrule
\(m_1\), eV                              &    \(0.0493_{-0.00046}^{+0.00041}\) &    \(0.0204_{-0.00035}^{+0.00042}\) &                               \(0\) \\
\(m_2\), eV                              &      \(0.05_{-0.00042}^{+0.00037}\) &     \(0.0221_{-0.00028}^{+0.0003}\) &     \(0.0086_{-0.00026}^{+0.0002}\) \\
\(m_3\), eV                              &                               \(0\) &    \(0.0542_{-0.00046}^{+0.00054}\) &    \(0.0502_{-0.00043}^{+0.00046}\) \\
\(\Sigma_i m_i\), eV                     &      \(0.0993_{-0.0009}^{+0.0008}\) &       \(0.0967_{-0.001}^{+0.0013}\) &      \(0.0588_{-0.0002}^{+0.0002}\) \\
\(\left| \langle m \rangle \right|\), eV &       \(0.0197_{-0.0031}^{+0.002}\) &      \(0.0181_{-0.0003}^{+0.0004}\) &   \(0.00144_{-0.00033}^{+0.00035}\) \\
\(\delta / \pi\)                         &            \(1.88_{-0.13}^{+0.37}\) &            \(1.44_{-0.01}^{+0.01}\) &                \(1 \pm \mathcal{O}(10^{-6})\) \\
\(\alpha_{21} / \pi\)                    &            \(0.91_{-0.09}^{+0.28}\) &            \(1.77_{-0.01}^{+0.01}\) &                               \(0\) \\
\(\alpha_{31} / \pi\)                    &                               \(0\) &            \(1.86_{-0.02}^{+0.02}\) &                \(1 \pm \mathcal{O}(10^{-5})\) \\
\midrule
\(N \sigma\)                             &                           \(0.431\) &                           \(0.649\) &                           \(0.563\) \\
\bottomrule
  \end{tabular}
  \caption{Best-fit values and 3\(\sigma\) ranges of the parameters and observables for the models discussed in sections~\ref{sec:models_inf} and~\ref{sec:models_S4_omega}}
  \label{tab:pheno}
\end{table}

\vfill
\clearpage

%%%%%%%%%%%%%%%%%%%%%%%%%%%%%%%
\section{Summary and Conclusions}
\label{sec:conclusions}
%%%%%%%%%%%%%%%%%%%%%%%%%%%%%%%

 We have investigated the possibility to obtain 
fermion mass hierarchies without fine-tuning 
in modular-invariant theories of flavour, 
which do not include flavons. In these theories, hierarchical 
fermion mass matrices may arise solely due to the proximity of 
the VEV of the modulus \(\tau\) to a point of residual symmetry.
In our analysis we have considered
theories with flavour symmetry 
described by a finite inhomogeneous or homogeneous 
modular group, \(\Gamma_N\) or \(\Gamma^{\prime}_N\),
with \(N\leq 5\). 
For \(N\leq 5\), the finite modular groups \(\Gamma_N\) 
are isomorphic to the permutation groups  
\(\Gamma_2\simeq S_3\), \(\Gamma_3\simeq A_4\), \(\Gamma_4 \simeq S_4\) and 
\(\Gamma_5\simeq A_5\), 
while the groups \(\Gamma^\prime_N\) are isomorphic to 
their double covers
\(S^\prime_3 \equiv S_3\), \(A^\prime_4 \equiv T^\prime\), 
\(S^\prime_4\) and \(A^\prime_5\). 

In the simplest class of such models considered by us, the
VEV of the modulus \(\tau\) is the only source of flavour 
symmetry breaking, such that no flavons are needed. 
Another appealing feature of the proposed framework is that the
VEV of \(\tau\) can also be the only source 
of CP symmetry breaking in the theory~\cite{Novichkov:2019sqv}.
There is no value of \(\tau\) which preserves 
the full modular symmetry. Nevertheless, at certain so-called symmetric 
points \(\tau = \tau_\text{sym}\) the modular group is only partially broken, with the unbroken generators giving rise to residual symmetries.
There are only three inequivalent symmetric points in the fundamental domain 
of the modular group~\cite{Novichkov:2018ovf}:
\(\tau_\text{sym} = i\), \(\tau_\text{sym} = \omega \equiv \exp(i\,2\pi/ 3)
= -\,1/2 + i\sqrt{3}/2\) (the `left cusp'),
and \(\tau_\text{sym} = i\infty\).
In these three points, the theories
based on \(\Gamma_N\) invariance 
have respectively 
\(\mathbb{Z}^{S}_2\),  \(\mathbb{Z}^{ST}_3\) and 
 \(\mathbb{Z}^{T}_N\) residual symmetries.
 In the case of the double cover groups \(\Gamma^\prime_N\),
there is an additional \(\mathbb{Z}_2^R\) symmetry
that is unbroken for any value of \(\tau\)~\cite{Novichkov:2020eep}, 
thus enlarging the residual 
symmetries \(\mathbb{Z}^{ST}_3\) and 
\(\mathbb{Z}^{T}_N\)  by the factor   
\(\mathbb{Z}_2^R\), while the \(\mathbb{Z}^{S}_2\)  symmetry is 
enlarged to a \(\mathbb{Z}^{S}_4\) one.
In each of the three symmetric points the standard 
\(\mathbb{Z}_2^{\rm CP}\) symmetry may also be conserved~\cite{Novichkov:2019sqv}. 

The indicated residual symmetries play a crucial role 
in our analysis. In \(\Gamma^{(\prime)}_N\)
modular invariant theories of flavour 
the fermion mass matrices are modular forms
of a given level \(N\). 
As we show, the mass matrices 
can be strongly constrained in the vicinity of points of residual symmetries 
by the properties of the respective modular forms. 
For each of the three symmetric points, we have developed the 
formalism which allows to determine the degree of suppression 
of the elements of the fermion mass matrices, and correspondingly, of 
their singular values -- the fermion masses --
in the vicinity of a given symmetric point.
More specifically, our analysis showed that,
if \(\epsilon\) parameterises the deviation of \(\tau\) from 
a given symmetric point, \(|\epsilon|\ll 1\), 
the degree of suppression is given by 
\(|\epsilon|^l\), where \(l\) is an integer and can take values
\begin{enumerate}[i)]
    \item \(l = 0,1,...,N-1\) in the case of \(\tau_\text{sym} = i\infty\),
    \item \(l = 0,1,2\) if  \(\tau_\text{sym} = \omega\), and
    \item \(l = 0,1\) when  \(\tau_\text{sym} = i\).
\end{enumerate}
These results show, in particular, that it is impossible 
to obtain the charged-lepton and quark
mass hierarchies in the vicinity of the symmetric point 
\(\tau_\text{sym} = i\) as a consequence only of the 
smallness of \(|\epsilon|\).
As we have proven, the specific value of the power \(l\) 
depends only on how the representations of the fermion fields in the 
mass term bilinear, denoted for brevity as \(\psi_i\) and \(\psi^c_j\), 
decompose under the considered residual symmetry group. 
We have derived the decompositions of the weighted irreducible 
representations of \(\Gamma^\prime_N\) (\(N\leq 5\)) 
under the three residual symmetry groups, i.e., 
the residual decompositions of the irreducible representations (irreps) 
of \(\Gamma_2' \simeq S_3\), \(\Gamma_3' \simeq A_4' = T'\), 
\(\Gamma_4' \simeq S_4' = SL(2,\mathbb{Z}_4)\), and 
\(\Gamma_5' \simeq A_5' = SL(2,\mathbb{Z}_5)\) 
(they are listed in Tables \ref{tab:S3_residual_reps}\,--\,\ref{tab:A5p_residual_reps} in Appendix~\ref{app:decomp}).
The results include also the case of irreps of
\(\Gamma_N\), since they represent a subset of the irreps of 
\(\Gamma_N'\).

 Having these results we proceeded to identify 
\(3\times 3\) hierarchical fermion mass matrices where the hierarchical 
pattern is a result of the proximity of the modulus to a point of 
residual symmetry and no massless fermions are present in the spectrum. 
We analysed bilinears of the type  
\(\psi^c_i \, M(\tau)_{ij}\, \psi_j\), \(M\) being the mass matrix,
and considered all possible 3-dimensional representations for 
the fields \(\psi_i\) and \(\psi^c_j\), \(i,j=1,2,3\). 
While the representations of these fields 
\(\mathbf{r}\) and \(\mathbf{r}^c\) 
are in general reducible, we focused on the case where 
the same weight is shared between the irreps into which they decompose.
The results of this analysis are given in 
Tables~\ref{tab:S3_patterns}\,--\,\ref{tab:A5p_patterns} of 
Appendix~\ref{app:patterns}. These tables summarise, 
for each of the levels \(N \leq 5\), the patterns of the three 
fermion masses which may arise 
in the vicinity of the two potentially viable symmetric points, 
\(\tau_\text{sym} = \omega\) and \(i \infty\), 
for all \((\mathbf{r},\mathbf{r}^c)\) pairs of 
3-dimensional representations and all weights \(k^{(c)}\).
We have found that it is only possible to obtain hierarchical spectra 
for a small list of representation pairs, the most promising of which 
are collected in Table~\ref{tab:good_patterns} 
and correspond to the patterns 
\((1,\epsilon,\epsilon^2)\), \((1,\epsilon,\epsilon^3)\) and 
\((1,\epsilon,\epsilon^4)\).

Using the developed formalism and the aforementioned results we have performed 
a scan searching for phenomenologically viable models of lepton flavour
where the charged-lepton mass hierarchies 
are a consequence of the described mechanism.
The charged-lepton masses are obtained from their Yukawa interactions,
while neutrino masses are generated either by the Weinberg operator
or within a type-I seesaw UV completion.
Aiming at minimal and predictive models, we have imposed a 
generalised CP symmetry on the models, enforcing
the reality of coupling 
constants~\cite{Novichkov:2019sqv}
and restricted the scan to models involving at most 8
constant parameters (including \(\re \tau\) and \(\im \tau\)).
Models producing a massless electron were rejected.
Out of the many models thus identified, we have selected 
only those which 
i) are phenomenologically viable in the regime of interest, and
ii) produce a charged-lepton spectrum which is not fine-tuned.

We have found two viable models and both are 
in the vicinity of~\(\tau = i \infty\).
The first is an \(A_5'\) model  
where the lepton doublets \(L\) and charged-lepton singlets \(E^c\) 
are different triplets of \(A_5'\).
For this model, the charged-lepton, Dirac-neutrino and \(N^c\) Majorana mass terms involve modular forms of weights 4, 5 and 4, respectively.
The neutrino masses are generated by 
the seesaw mechanism with the gauge-singlet neutrino fields \(N^c\) 
furnishing a doublet of \(A_5'\).
The predicted charged-lepton mass pattern is 
\((m_{\tau}, m_{\mu}, m_e) \sim (1, \epsilon, \epsilon^4)\).
The best description of the input data 
on the charged-lepton and neutrino masses and mixing 
(\(N\sigma = 0.43\)) was found to be obtained 
for \(\re \tau = -0.47\), \(\im \tau = 3.11\). 

The second viable model found by us is based on \(S_4'\)
modular symmetry. 
In this model the charged-lepton, Dirac-neutrino and \(N^c\) Majorana 
mass terms involve modular forms of weights 4, 3 and 2, respectively.
The charged-lepton mass pattern 
is predicted to be 
\(\left( m_{\tau}, m_{\mu}, m_e \right) \sim (1, \epsilon, \epsilon^3)\).
The viable region in the \(\tau\) plane is centred around \(\tau = 2.65\,i\).

 Both the \(A_5'\) and  \(S_4'\) viable models were found to require a certain 
amount of fine-tuning when describing the neutrino masses and mixing.
The presence of fine-tuning in the neutrino sector is explained 
by the necessity to introduce large corrections to the symmetric-limit 
PMNS matrix.
Addressing the problem of fine-tuning in the neutrino sector,
 we have found that a modular-symmetric model of 
lepton flavour with hierarchical charged-lepton masses is expected 
to be free of fine-tuning%
\footnote{In other words, it is possible to have a PMNS matrix which
is close to the observed one even in the symmetric limit, i.e., 
such that either none of its entries vanish, or only 
the (13) entry vanishes as \(\epsilon \rightarrow 0\).
} 
if it satisfies at least one of four conditions 
(see section~\ref{sec:natural}).
Two of the conditions were formulated earlier in 
Ref.~\cite{Reyimuaji:2018xvs} for arbitrary 
flavour symmetry groups.

We have constructed  a viable model based on \(S_4'\) 
modular symmetry  in the vicinity of~\(\tau = \omega\), 
which is free of fine-tuning in both the charged-lepton and 
neutrino sectors. It has altogether nine parameters. 
The neutrino masses are generated via the seesaw mechanism.
The charged-lepton mass pattern  is predicted to be 
\(\left( m_{\tau}, m_{\mu}, m_e \right) \sim (1, \epsilon, \epsilon^2)\).
The model predicts, in particular, \(\delta \simeq \pi\) and 
\(m_1 \simeq 0\). We have found also other viable non--fine-tuned \(S_4'\) models which, however, predict \(\delta \simeq 0\). 

 For the three lepton flavour models constructed,
we collect in Table~\ref{tab:pheno}
 the best-fit values and \(3\sigma\) allowed ranges of i) \(\re \tau\), \(\im \tau\) and the superpotential parameters,
 of ii) the charged-lepton masses and neutrino mass and mixing observables 
used as input in the statistical analysis of the models, and
 of iii)
 the predicted lightest neutrino mass, the Dirac and Majorana CPV phases, the sum of the neutrino masses and effective neutrinoless double beta decay Majorana mass.

 The results obtained in the present article show, in particular,
that the requirement of absence of fine-tuning in both the charged 
lepton and neutrino sectors in lepton flavour models based on modular 
invariance is remarkably restrictive. 
It is hoped that 
using this requirement it might be possible to identify 
not more than a few, if not just one,
modular-invariant models 
providing a simultaneous, viable and appealing solution 
to both the lepton and quark flavour problems.

\vskip 1cm
\noindent 
{\it Note Added.}  While this work was in its conclusion,
 Ref.~\cite{Feruglio:2021dte} appeared on the arXiv in which 
the authors investigated the possibility 
to generate the charged-lepton mass hierarchy in the vicinity of the 
symmetric point  \(\tau_\text{sym} = i\). 
The two models presented in  Ref.~\cite{Feruglio:2021dte} 
are restricted to level \(N=3\). In these scenarios, the electron 
is massless by construction, \(m_e = 0\), and \(m_e\neq 0\) is expected 
to arise either through SUSY breaking or from a dim-6 operator. 
The ratios \(m_e/m_\tau\) and \(m_\mu/m_\tau\) are then associated to 
independent parameters and not to different powers of the same 
expansion parameter, as is the case of our work.

%%%%%%%%%%%%%%%%%%%%%%%
\section*{Acknowledgements}
%%%%%%%%%%%%%%%%%%%%%%%
%
  This project has received funding/support from the European Union's Horizon 2020 research and innovation programme under the Marie Skłodowska-Curie grant agreement No.~860881-HIDDeN.
  This work was supported in part
  by the INFN program on Theoretical Astroparticle Physics (P.P.N. and S.T.P.)
  and by the  World Premier International Research Center
  Initiative (WPI Initiative, MEXT), Japan (S.T.P.).
  The work of J.T.P.~was supported by
  Fundação para a Ciência e a Tecnologia (FCT, Portugal) through the projects
  PTDC/FIS-PAR/29436/2017, 
  CERN/FIS-PAR/0004/2019, CERN/FIS-PAR/0008/2019, and
  CFTP-FCT Unit 777 (namely UIDB/00777/2020 and UIDP/00777/2020),
  which are partially funded through POCTI (FEDER), COMPETE, QREN and EU.

\vfill
\pagebreak

%%%%%%%%%%%%%%%%%%%%%%%
\appendix
%%%%%%%%%%%%%%%%%%%%%%%

%%%%%%%%%%%%%%%%%%%%%%%
\section{Residual group decompositions}
\label{app:decomp}
%%%%%%%%%%%%%%%%%%%%%%%
%
The multiplets of \(\Gamma_N'\) are `weighted', i.e.~are described by a pair \((\mathbf{r}, k)\).%
\footnote{As in~\cite{Novichkov:2020eep}, we denote with a hat representations \(\mathbf{r}\) for which \(\rho_\mathbf{r}(R) = -\id\).}
At a symmetric point these multiplets decompose into 1-dimensional representations of the corresponding residual symmetry group. In this appendix we present the decompositions of \(\Gamma_N'\) multiplets (\(N \leq 5\)) under the three residual groups of interest (Tables~\ref{tab:S3_residual_reps}\,--\,\ref{tab:A5p_residual_reps}). As seen in section~\ref{sec:residual}, these are \(\mathbb{Z}_4^S\), \(\mathbb{Z}_3^{ST} \times \mathbb{Z}_2^R\) and \(\mathbb{Z}_N^T\times \mathbb{Z}_2^R\).

Before proceeding, let us comment on the \(\mathbb{Z}_2^R\) factors in \(\mathbb{Z}_3^{ST} \times \mathbb{Z}_2^R\) and in  \(\mathbb{Z}_N^T\times \mathbb{Z}_2^R\). While kept as part of the residual symmetry group definition in this appendix, they have been omitted in the main text of section~\ref{sec:decomp}.
To understand why they can be ignored without loss of generality, note that a direct product \(\mathbb{Z}_n \times \mathbb{Z}_2 \equiv \left\langle a, b \, \vert a^n = b^2 = 1, \, ab = ba \right\rangle\) has \(2n\) irreps \(\mathbf{1}_k^{\pm}\), \(k = 0, \ldots, n-1\), which are simply given as products of the \(\mathbb{Z}_n\) and \(\mathbb{Z}_2\) irreps:
\begin{equation}
  \label{eq:ZnZ2_reps}
  \mathbf{1}_k^{\pm} \, : \quad \rho(a) = \exp \left( 2\pi i \frac{k}{n} \right) \,, \quad \rho(b) = \pm 1 \, .
\end{equation}
In this notation, \(1_0^+\) is the trivial irrep. The representation under \(\mathbb{Z}_2\) is just a sign and does not affect the reality/complexity of a representation. Hence real irreps are \(1_0^+\), \(1_0^-\) and, for even \(n\), \(1_{n/2}^+\), \(1_{n/2}^-\) (one also has \((\mathbf{1}_k^{\pm})^{*} = \mathbf{1}_{n-k}^{\pm}\)).
Since \(M(\tau)\) in the bilinear of eq.~\eqref{eq:bilinear} is a function of \(\tau\) alone, it is \(R\)-even. The fields \(\psi\) and \(\psi^c\) are then constrained to carry the same \(R\)-parity, i.e.~transform with the same sign under \(\mathbb{Z}_2^R\). Fields in unhatted representations \(\mathbf{r}\) -- for which \(\rho_\mathbf{r}(R) = \id\) -- are even (odd) under \(\mathbb{Z}_2^R\) if \(k\) is even (odd), while the opposite happens for hatted representations. Keeping this in mind, one can omit the \(\mathbb{Z}_2^R\) factor and ignore the superscript signs in the following tables.

Finally, notice that a \(\mathbb{Z}_2^R\) factor is hidden in the residual \(\mathbb{Z}_4^S\), as \(S^2 = R\). Fields transforming under \(\mathbb{Z}_4^S\) as \(1_0\) or \(1_2\) are \(R\)-even while fields transforming as \(\mathbf{1}_1\) or \(\mathbf{1}_3\) are \(R\)-odd. Requiring that \(\psi\) and \(\psi^c\) carry the same \(R\)-parity implies that one effectively works with \(\mathbb{Z}_4^S / \mathbb{Z}_2^R \simeq \mathbb{Z}_2\), which is why it is generic to consider \(\tilde\rho^c_i \tilde\rho_j = \pm 1\) in section~\ref{sec:theory_S}.

\vfill
\begin{table}[ht!]
  \centering
  \renewcommand{\arraystretch}{1.7}
  \begin{tabular}{llll}
  \toprule  
\(\mathbf{r}\qquad\) & \(\mathbb{Z}_4^S \,(\tau=i)\qquad\)
& \(\mathbb{Z}_3^{ST} \times  \mathbb{Z}_2^R \,(\tau=\omega)\qquad\)
& \(\mathbb{Z}_2^{T} \times  \mathbb{Z}_2^R \,(\tau=i\infty)\qquad\)\\
\midrule
%%%%%%%%%%%%
\(\mathbf{1}\) &
 \(\mathbf{1}_k\) &
 \(\mathbf{1}_k^\pm\) &
 \({1}_0^\pm\) \\
%%%%%%%%%%%%
\(\mathbf{1'}\) &
 \(\mathbf{1}_{k+2}\) &
 \(\mathbf{1}_k^\pm\) &
 \({1}_1^\pm\) \\
%%%%%%%%%%%%
\(\mathbf{2}\) &
 \(\mathbf{1}_k \oplus \mathbf{1}_{k+2}\) &
 \(\mathbf{1}_{k-1}^\pm \oplus \mathbf{1}_{k+1}^\pm\) &
 \({1}_0^\pm \oplus {1}_1^\pm\) \\
%%%%%%%%%%%%
\bottomrule
  \end{tabular}
  \caption{
    Decompositions of `weighted' \((\mathbf{r}, k)\) multiplets of \(\Gamma_2'\simeq S_3\) under the residual symmetry groups.
    Irrep subscripts should be understood modulo \(n\), where \(n = 4, 3\) in the first and second columns, respectively.
    Upper (lower) signs correspond to even (odd) values of \(k\).
  }
  \label{tab:S3_residual_reps}
\end{table}

\begin{table}[ht!]
  \centering
  \renewcommand{\arraystretch}{1.7}
  \begin{tabular}{llll}
  \toprule
\(\mathbf{r}\qquad\) & \(\mathbb{Z}_4^S \,(\tau=i)\qquad\)
& \(\mathbb{Z}_3^{ST} \times  \mathbb{Z}_2^R \,(\tau=\omega)\qquad\)
& \(\mathbb{Z}_3^{T} \times  \mathbb{Z}_2^R \,(\tau=i\infty)\qquad\)\\
\midrule
%%%%%%%%%%%%
\(\mathbf{1}\) &
 \(\mathbf{1}_k\) &
 \(\mathbf{1}_{k}^\pm\) &
 \(1_0^\pm\) \\
%%%%%%%%%%%%
\(\mathbf{1'}\) &
 \(\mathbf{1}_k\) &
 \(\mathbf{1}_{k+1}^\pm\) &
 \(\mathbf{1}_1^\pm\) \\
%%%%%%%%%%%%
\(\mathbf{1''}\) &
 \(\mathbf{1}_k\) &
 \(\mathbf{1}_{k+2}^\pm\) &
 \(\mathbf{1}_2^\pm\) \\
%%%%%%%%%%%%
\(\mathbf{\hat{2}}\) &
 \(\mathbf{1}_{k+1} \oplus \mathbf{1}_{k+3}\) &
 \(\mathbf{1}_{k}^\mp \oplus \mathbf{1}_{k+1}^\mp\) &
 \(1_0^\mp \oplus \mathbf{1}_1^\mp\) \\
%%%%%%%%%%%%
\(\mathbf{\hat{2}'}\) &
 \(\mathbf{1}_{k+1} \oplus \mathbf{1}_{k+3}\) &
 \(\mathbf{1}_{k+1}^\mp \oplus \mathbf{1}_{k+2}^\mp\) &
 \(\mathbf{1}_1^\mp \oplus \mathbf{1}_2^\mp\) \\
%%%%%%%%%%%%
\(\mathbf{\hat{2}''}\) &
 \(\mathbf{1}_{k+1} \oplus \mathbf{1}_{k+3}\) &
 \(\mathbf{1}_{k}^\mp \oplus \mathbf{1}_{k+2}^\mp\) &
 \(1_0^\mp \oplus \mathbf{1}_2^\mp\) \\
%%%%%%%%%%%%
\(\mathbf{3}\) &
 \(\mathbf{1}_{k} \oplus \mathbf{1}_{k+2} \oplus \mathbf{1}_{k+2}\) &
 \(\mathbf{1}_{k}^\pm \oplus \mathbf{1}_{k+1}^\pm \oplus \mathbf{1}_{k+2}^\pm\) &
 \({1}_0^\pm \oplus \mathbf{1}_{1}^\pm \oplus \mathbf{1}_{2}^\pm\) \\
%%%%%%%%%%%%
\bottomrule
  \end{tabular}
  \caption{  
    Decompositions of `weighted' \((\mathbf{r}, k)\) multiplets of \(\Gamma_3'\simeq A_4' = T'\) under the residual symmetry groups.
    Irrep subscripts should be understood modulo \(n\), where \(n = 4, 3\) in the first and second columns, respectively.
    Upper (lower) signs correspond to even (odd) values of \(k\).
}
  \label{tab:A4p_residual_reps}
\end{table}

\begin{table}[ht!]
  \centering
  \renewcommand{\arraystretch}{1.7}
  \begin{tabular}{llll}
  \toprule
\(\mathbf{r}\qquad\) & \(\mathbb{Z}_4^S \,(\tau=i)\qquad\)
& \(\mathbb{Z}_3^{ST} \times  \mathbb{Z}_2^R \,(\tau=\omega)\qquad\)
& \(\mathbb{Z}_4^{T} \times  \mathbb{Z}_2^R \,(\tau=i\infty)\qquad\)\\
\midrule
%%%%%%%%%%%%
\(\mathbf{1}\) &
 \(\mathbf{1}_k\) &
 \(\mathbf{1}_k^{\pm}\) &
 \(1_0^{\pm}\) \\
%%%%%%%%%%%%
\(\mathbf{\hat{1}}\) &
 \(\mathbf{1}_{k+1}\) & 
 \(\mathbf{1}_k^{\mp}\) &
 \(\mathbf{1}_3^{\mp}\) \\
%%%%%%%%%%%%
\(\mathbf{1}'\) & 
 \(\mathbf{1}_{k+2}\) &
 \(\mathbf{1}_k^{\pm}\) &
 \(1_2^{\pm}\) \\
%%%%%%%%%%%%
\(\mathbf{\hat{1}}'\) & 
 \(\mathbf{1}_{k+3}\) & 
 \(\mathbf{1}_k^{\mp}\) & 
 \(\mathbf{1}_1^{\mp}\) \\
%%%%%%%%%%%%
\(\mathbf{2}\) &
 \(\mathbf{1}_{k+2} \oplus \mathbf{1}_k\) &
 \(\mathbf{1}_{k+1}^{\pm} \oplus \mathbf{1}_{k+2}^{\pm}\) &
 \(1_0^{\pm} \oplus 1_2^{\pm}\) \\
%%%%%%%%%%%%
\(\mathbf{\hat{2}}\) &
 \(\mathbf{1}_{k+1} \oplus \mathbf{1}_{k+3}\) & 
 \(\mathbf{1}_{k+1}^{\mp} \oplus \mathbf{1}_{k+2}^{\mp}\) &
 \(\mathbf{1}_1^{\mp} \oplus \mathbf{1}_3^{\mp}\) \\
%%%%%%%%%%%%
\(\mathbf{3}\) &
 \(\mathbf{1}_{k+2} \oplus \mathbf{1}_k \oplus \mathbf{1}_k\) &
 \(\mathbf{1}_k^{\pm} \oplus \mathbf{1}_{k+1}^{\pm} \oplus \mathbf{1}_{k+2}^{\pm}\) &
 \(\mathbf{1}_1^{\pm} \oplus 1_2^{\pm} \oplus \mathbf{1}_3^{\pm}\) \\
%%%%%%%%%%%%
\(\mathbf{\hat{3}}\) &
 \(\mathbf{1}_{k+1} \oplus \mathbf{1}_{k+1} \oplus \mathbf{1}_{k+3}\) &
 \(\mathbf{1}_k^{\mp} \oplus \mathbf{1}_{k+1}^{\mp} \oplus \mathbf{1}_{k+2}^{\mp}\) &
 \(1_0^{\mp} \oplus \mathbf{1}_1^{\mp} \oplus 1_2^{\mp}\) \\
%%%%%%%%%%%%
\(\mathbf{3}'\) &
 \(\mathbf{1}_{k+2} \oplus \mathbf{1}_{k+2} \oplus \mathbf{1}_k\) & 
 \(\mathbf{1}_k^{\pm} \oplus \mathbf{1}_{k+1}^{\pm} \oplus \mathbf{1}_{k+2}^{\pm}\) &
 \(1_0^{\pm} \oplus \mathbf{1}_1^{\pm} \oplus \mathbf{1}_3^{\pm}\) \\
%%%%%%%%%%%%
\(\mathbf{\hat{3}}'\) &
 \(\mathbf{1}_{k+1} \oplus \mathbf{1}_{k+3} \oplus \mathbf{1}_{k+3}\) &
 \(\mathbf{1}_k^{\mp} \oplus \mathbf{1}_{k+1}^{\mp} \oplus \mathbf{1}_{k+2}^{\mp}\) &
 \(1_0^{\mp} \oplus 1_2^{\mp} \oplus \mathbf{1}_3^{\mp}\) \\
%%%%%%%%%%%%
\bottomrule
  \end{tabular}
  \caption{
    Decompositions of `weighted' \((\mathbf{r}, k)\) multiplets of \(\Gamma_4'\simeq S_4'=SL(2,\mathbb{Z}_4)\) under the residual symmetry groups.
    Irrep subscripts should be understood modulo \(n\), where \(n = 4, 3\) in the first and second columns, respectively.
    Upper (lower) signs correspond to even (odd) values of \(k\).
  }
  \label{tab:S4p_residual_reps}
\end{table}

\vfill
\clearpage

\newgeometry{left=4cm,right=4cm,top=2.5cm,bottom=2.5cm}
\begin{landscape}
\thispagestyle{empty}

\begin{table}[ht!]
  \centering
  \renewcommand{\arraystretch}{1.7}
  \begin{tabular}{llll}
  \toprule
\(\mathbf{r}\) & \(\mathbb{Z}_4^S \,(\tau=i)\)
& \(\mathbb{Z}_3^{ST} \times  \mathbb{Z}_2^R \,(\tau=\omega)\)
& \(\mathbb{Z}_5^{T} \times  \mathbb{Z}_2^R \,(\tau=i\infty)\)\\
\midrule
%%%%%%%%%%%%
\(\mathbf{1}\) &
 \(\mathbf{1}_k\) &
 \(\mathbf{1}_{k}^\pm\) &
 \(1_0^\pm\) 
 \\
%%%%%%%%%%%%
\(\mathbf{\hat{2}}\) &
 \(\mathbf{1}_{k+1} \oplus \mathbf{1}_{k+3}\) &
 \(\mathbf{1}_{k+1}^\mp \oplus \mathbf{1}_{k+2}^\mp\) &
 \(\mathbf{1}_2^\mp \oplus \mathbf{1}_3^\mp\) 
 \\
%%%%%%%%%%%%
\(\mathbf{\hat{2}'}\) &
 \(\mathbf{1}_{k+1} \oplus \mathbf{1}_{k+3}\) &
 \(\mathbf{1}_{k+1}^\mp \oplus \mathbf{1}_{k+2}^\mp\) &
 \(\mathbf{1}_1^\mp \oplus \mathbf{1}_4^\mp\) 
 \\
%%%%%%%%%%%%
\(\mathbf{3}\) &
 \(\mathbf{1}_{k} \oplus \mathbf{1}_{k+2} \oplus \mathbf{1}_{k+2}\) &
 \(\mathbf{1}_{k}^\pm \oplus \mathbf{1}_{k+1}^\pm \oplus \mathbf{1}_{k+2}^\pm\) &
 \({1}_0^\pm \oplus \mathbf{1}_{1}^\pm \oplus \mathbf{1}_{4}^\pm\) 
 \\
%%%%%%%%%%%%
\(\mathbf{3'}\) &
 \(\mathbf{1}_{k} \oplus \mathbf{1}_{k+2} \oplus \mathbf{1}_{k+2}\) &
 \(\mathbf{1}_{k}^\pm \oplus \mathbf{1}_{k+1}^\pm \oplus \mathbf{1}_{k+2}^\pm\) &
 \({1}_0^\pm \oplus \mathbf{1}_{2}^\pm \oplus \mathbf{1}_{3}^\pm\) 
 \\
%%%%%%%%%%%%
\(\mathbf{4}\) &
 \(\mathbf{1}_{k} \oplus \mathbf{1}_{k} \oplus \mathbf{1}_{k+2} \oplus \mathbf{1}_{k+2}\) &
 \(\mathbf{1}_{k}^\pm \oplus \mathbf{1}_{k}^\pm \oplus \mathbf{1}_{k+1}^\pm \oplus \mathbf{1}_{k+2}^\pm\) &
 \(\mathbf{1}_{1}^\pm \oplus \mathbf{1}_{2}^\pm \oplus \mathbf{1}_{3}^\pm \oplus \mathbf{1}_{4}^\pm\) 
 \\
%%%%%%%%%%%%
\(\mathbf{\hat{4}}\) &
 \(\mathbf{1}_{k+1} \oplus \mathbf{1}_{k+1} \oplus \mathbf{1}_{k+3} \oplus \mathbf{1}_{k+3}\) &
 \(\mathbf{1}_{k}^\mp \oplus
 \mathbf{1}_{k}^\mp \oplus
 \mathbf{1}_{k+1}^\mp \oplus \mathbf{1}_{k+2}^\mp\) &
 \(\mathbf{1}_{1}^\mp \oplus \mathbf{1}_{2}^\mp \oplus \mathbf{1}_{3}^\mp \oplus \mathbf{1}_{4}^\mp\) 
 \\
%%%%%%%%%%%%
\(\mathbf{5}\) &
 \(\mathbf{1}_{k} \oplus \mathbf{1}_{k} \oplus \mathbf{1}_{k} \oplus \mathbf{1}_{k+2} \oplus \mathbf{1}_{k+2}\) &
 \(\mathbf{1}_{k}^\pm \oplus \mathbf{1}_{k+1}^\pm \oplus \mathbf{1}_{k+1}^\pm \oplus \mathbf{1}_{k+2}^\pm \oplus \mathbf{1}_{k+2}^\pm\) &
 \({1}_0^\pm \oplus \mathbf{1}_{1}^\pm \oplus \mathbf{1}_{2}^\pm \oplus \mathbf{1}_{3}^\pm \oplus \mathbf{1}_{4}^\pm\)
 \\
%%%%%%%%%%%%
\(\mathbf{\hat{6}}\) &
 \(\mathbf{1}_{k+1} \oplus \mathbf{1}_{k+1} \oplus \mathbf{1}_{k+1} \oplus \mathbf{1}_{k+3}\oplus \mathbf{1}_{k+3} \oplus \mathbf{1}_{k+3}\) &
 \(\mathbf{1}_{k}^\mp \oplus
 \mathbf{1}_{k}^\mp \oplus
 \mathbf{1}_{k+1}^\mp \oplus
 \mathbf{1}_{k+1}^\mp \oplus
 \mathbf{1}_{k+2}^\mp \oplus \mathbf{1}_{k+2}^\mp\) &
 \({1}_0^\mp \oplus {1}_0^\mp \oplus \mathbf{1}_{1}^\mp \oplus \mathbf{1}_{2}^\mp \oplus \mathbf{1}_{3}^\mp \oplus \mathbf{1}_{4}^\mp\)
 \\
%%%%%%%%%%%%
\bottomrule
  \end{tabular}
  \caption{
    Decompositions of `weighted' \((\mathbf{r}, k)\) multiplets of \(\Gamma_5'\simeq A_5'=SL(2,\mathbb{Z}_5)\) under the residual symmetry groups.
    Irrep subscripts should be understood modulo \(n\), where \(n = 4, 3\) in the first and second columns, respectively.
    Upper (lower) signs correspond to even (odd) values of \(k\).
  }
  \label{tab:A5p_residual_reps}
\end{table}

\vfill
\begin{center}
    \thepage
\end{center}
\end{landscape}
\restoregeometry

\vfill
\clearpage

%%%%%%%%%%%%%%%%%%%%%%%
\section{Possible hierarchical patterns}
\label{app:patterns}
%%%%%%%%%%%%%%%%%%%%%%%
%
In this appendix we list the hierarchical patterns which may arise in the vicinities of the two symmetric points of interest (see main text). We consider in turn the finite modular groups \(\Gamma_2' \simeq S_3\), \(\Gamma_3' \simeq A_4' = T'\), \(\Gamma_4' \simeq S_4' = SL(2,\mathbb{Z}_4)\), and \(\Gamma_5' \simeq A_5' = SL(2,\mathbb{Z}_5)\) (Tables~\ref{tab:S3_patterns}\,--\,\ref{tab:A5p_patterns}). We have focused on 3-dimensional (possibly reducible) representations \((\mathbf{r},\mathbf{r}^c)\) entering the bilinear \eqref{eq:bilinear}. 
Dependence on the weights \(k^{(c)}\) may only arise for \(\tau \sim \omega\) and through the combination \(K = k+k^c\), modulo 3.
One can see from Tables~\ref{tab:S3_residual_reps}\,--\,\ref{tab:A5p_residual_reps} that if one of the 3d multiplets (say \(\psi\)) entering the bilinear is not a sum of 3 singlets, then its decomposition under the \(\mathbb{Z}^{ST}_3\) residual symmetry includes all possible singlets, \(1_0\), \(\mathbf{1}_1\), and \(\mathbf{1}_2\), independently of the weight \(k\).
In such cases, the hierarchies are independent of weights since a change in \(k^c\) can be absorbed by a change in \(k\) in their sum.

Note that for \(N=2\) the residual symmetry group at \(\tau_\text{sym}= i \infty\) is \(\mathbb{Z}_2^T\). Mass matrix entries are then expected to be either \(\mathcal{O}(1)\) or \(\mathcal{O}(\epsilon)\) and, as was the case for \(\tau \simeq i\), one cannot obtain the sought-after hierarchical patterns from the smallness of \(\epsilon\) alone. As such, only \(\tau \simeq \omega\) is considered in Table~\ref{tab:S3_patterns}.

\renewcommand{\arraystretch}{1.2}
\begin{longtable}{ccccc}
\caption{
    Leading-order mass spectra patterns of bilinears \(\psi^c\psi\) in the vicinity of the symmetric point \(\omega\), for 3d multiplets \(\psi \sim (\mathbf{r},k)\) and \(\psi^c \sim (\mathbf{r}^c,k^c)\) of the finite modular group \(\Gamma_2'\simeq S_3\). Spectra are insensitive to transposition, i.e.~to the exchange \(\psi\leftrightarrow \psi^c\). Congruence relations for \(k+k^c\) are modulo 3.
}
\label{tab:S3_patterns} \\
  \toprule
\multirow{2}{*}{\(\quad\mathbf{r}\quad\)} &\multirow{2}{*}{\(\quad\mathbf{r}^c\quad\)} & & \(\quad\tau\simeq\omega\quad\) & \\
 & & \(\,\,k+k^c \equiv 0 \,\,\)& \(\,\,k+k^c \equiv 1\,\, \) & \(\,\,k+k^c \equiv 2\,\, \)\\
\midrule \endfirsthead
\caption{(cont.)}\\
  \toprule
\multirow{2}{*}{\(\quad\mathbf{r}\quad\)} &\multirow{2}{*}{\(\quad\mathbf{r}^c\quad\)} & & \(\quad\tau\simeq\omega\quad\) & \\
 & & \(\,\,k+k^c \equiv 0 \,\,\)& \(\,\,k+k^c \equiv 1\,\, \) & \(\,\,k+k^c \equiv 2\,\, \)\\
\midrule \endhead
\bottomrule  \endfoot
%%%%%%%%%%%%
\(\mathbf{2} \oplus \mathbf{1}\) &
\(\mathbf{2} \oplus \mathbf{1}\) &
 \((1,1,1)\) &
 \((1,1,1)\) &
 \((1,1,1)\) \\ 
%%%%%%%%%%%%
\(\mathbf{2} \oplus \mathbf{1}\) &
\(\mathbf{2} \oplus \mathbf{1'}\) &
 \((1,1,1)\) &
 \((1,1,1)\) &
 \((1,1,1)\) \\ 
%%%%%%%%%%%%
\(\mathbf{2} \oplus \mathbf{1'}\) &
\(\mathbf{2} \oplus \mathbf{1'}\) &
 \((1,1,1)\) &
 \((1,1,1)\) &
 \((1,1,1)\) \\ 
%%%%%%%%%%%%
\(\mathbf{2} \oplus \mathbf{1}\) &
\(\mathbf{1'} \oplus \mathbf{1} \oplus \mathbf{1}\) &
 \((1,\epsilon,\epsilon^2)\) &
 \((1,\epsilon,\epsilon^2)\) &
 \((1,\epsilon,\epsilon^2)\) \\ 
%%%%%%%%%%%%
\(\mathbf{2} \oplus \mathbf{1}\) &
\(\mathbf{1'} \oplus \mathbf{1'} \oplus \mathbf{1}\) &
 \((1,\epsilon,\epsilon^2)\) &
 \((1,\epsilon,\epsilon^2)\) &
 \((1,\epsilon,\epsilon^2)\) \\ 
%%%%%%%%%%%%
\(\mathbf{2} \oplus \mathbf{1'}\) &
\(\mathbf{1'} \oplus \mathbf{1} \oplus \mathbf{1}\) &
 \((1,\epsilon,\epsilon^2)\) &
 \((1,\epsilon,\epsilon^2)\) &
 \((1,\epsilon,\epsilon^2)\) \\ 
%%%%%%%%%%%%
\(\mathbf{2} \oplus \mathbf{1'}\) &
\(\mathbf{1'} \oplus \mathbf{1'} \oplus \mathbf{1}\) &
 \((1,\epsilon,\epsilon^2)\) &
 \((1,\epsilon,\epsilon^2)\) &
 \((1,\epsilon,\epsilon^2)\) \\ 
%%%%%%%%%%%%
\(\mathbf{2} \oplus \mathbf{1}\) &
\(\mathbf{1} \oplus \mathbf{1} \oplus \mathbf{1}\) &
 \((1,\epsilon,\epsilon^2)\) &
 \((1,\epsilon,\epsilon^2)\) &
 \((1,\epsilon,\epsilon^2)\) \\
\end{longtable}

\renewcommand{\arraystretch}{1.2}
\begin{longtable}{cccccc}
\caption{
    Leading-order mass spectra patterns of bilinears \(\psi^c\psi\) in the vicinity of the symmetric points \(\omega\) and \(i \infty\), for 3d multiplets \(\psi \sim (\mathbf{r},k)\) and \(\psi^c \sim (\mathbf{r}^c,k^c)\) of the finite modular group \(\Gamma_3'\simeq A_4' = T'\). Spectra are insensitive to transposition, i.e.~to the exchange \(\psi\leftrightarrow \psi^c\). Congruence relations for \(k+k^c\) are modulo 3.
}
\label{tab:A4p_patterns} \\
  \toprule
\multirow{2}{*}{\(\quad\mathbf{r}\quad\)} &\multirow{2}{*}{\(\quad\mathbf{r}^c\quad\)} & & \(\quad\tau\simeq\omega\quad\) & & \multirow{2}{*}{\(\quad\tau\simeq i\infty\quad\)} \\
 & & \(\,\,k+k^c \equiv 0 \,\,\)& \(\,\,k+k^c \equiv 1\,\, \) & \(\,\,k+k^c \equiv 2\,\, \) &\\
\midrule \endfirsthead
\caption{(cont.)}\\
  \toprule
\multirow{2}{*}{\(\quad\mathbf{r}\quad\)} &\multirow{2}{*}{\(\quad\mathbf{r}^c\quad\)} & & \(\quad\tau\simeq\omega\quad\) & & \multirow{2}{*}{\(\quad\tau\simeq i\infty\quad\)} \\
 & & \(\,\,k+k^c \equiv 0 \,\,\)& \(\,\,k+k^c \equiv 1\,\, \) & \(\,\,k+k^c \equiv 2\,\, \) &\\
\midrule \endhead
\bottomrule  \endfoot
%%%%%%%%%%%%
\(\mathbf{3}\) &
\(\mathbf{3}\) &
 \((1,1,1)\) &
 \((1,1,1)\) &
 \((1,1,1)\) &
 \((1,1,1)\) \\ 
%%%%%%%%%%%%
\(\mathbf{3}\) &
\(\mathbf{1''} \oplus \mathbf{1'} \oplus \mathbf{1}\) &
 \((1,1,1)\) &
 \((1,1,1)\) &
 \((1,1,1)\) &
 \((1,1,1)\) \\ 
%%%%%%%%%%%%
\(\mathbf{3}\) &
\(\mathbf{1'} \oplus \mathbf{1} \oplus \mathbf{1}\) &
 \((1,1,\epsilon^2)\) &
 \((1,1,\epsilon^2)\) &
 \((1,1,\epsilon^2)\) &
 \((1,1,\epsilon^2)\) \\ 
%%%%%%%%%%%%
\(\mathbf{3}\) &
\(\mathbf{1''} \oplus \mathbf{1} \oplus \mathbf{1}\) &
 \((1,1,\epsilon)\) &
 \((1,1,\epsilon)\) &
 \((1,1,\epsilon)\) &
 \((1,1,\epsilon)\) \\ 
%%%%%%%%%%%%
\(\mathbf{3}\) &
\(\mathbf{1'} \oplus \mathbf{1'} \oplus \mathbf{1}\) &
 \((1,1,\epsilon)\) &
 \((1,1,\epsilon)\) &
 \((1,1,\epsilon)\) &
 \((1,1,\epsilon)\) \\ 
%%%%%%%%%%%%
\(\mathbf{3}\) &
\(\mathbf{1''} \oplus \mathbf{1''} \oplus \mathbf{1}\) &
 \((1,1,\epsilon^2)\) &
 \((1,1,\epsilon^2)\) &
 \((1,1,\epsilon^2)\) &
 \((1,1,\epsilon^2)\) \\ 
%%%%%%%%%%%%
\(\mathbf{3}\) &
\(\mathbf{1''} \oplus \mathbf{1'} \oplus \mathbf{1'}\) &
 \((1,1,\epsilon^2)\) &
 \((1,1,\epsilon^2)\) &
 \((1,1,\epsilon^2)\) &
 \((1,1,\epsilon^2)\) \\ 
%%%%%%%%%%%%
\(\mathbf{3}\) &
\(\mathbf{1''} \oplus \mathbf{1''} \oplus \mathbf{1'}\) &
 \((1,1,\epsilon)\) &
 \((1,1,\epsilon)\) &
 \((1,1,\epsilon)\) &
 \((1,1,\epsilon)\) \\
\end{longtable}

\renewcommand{\arraystretch}{1.2}
\begin{longtable}{cccccc}
\caption{
    Leading-order mass spectra patterns of bilinears \(\psi^c\psi\) in the vicinity of the symmetric points \(\omega\) and \(i \infty\), for 3d multiplets \(\psi \sim (\mathbf{r},k)\) and \(\psi^c \sim (\mathbf{r}^c,k^c)\) of the finite modular group \(\Gamma_4'\simeq S_4' = SL(2,\mathbb{Z}_4)\). Spectra are insensitive to transposition, i.e.~to the exchange \(\psi\leftrightarrow \psi^c\). Congruence relations for \(k+k^c\) are modulo 3.
}
\label{tab:S4p_patterns} \\
  \toprule
\multirow{2}{*}{\(\quad\mathbf{r}\quad\)} &\multirow{2}{*}{\(\quad\mathbf{r}^c\quad\)} & & \(\quad\tau\simeq\omega\quad\) & & \multirow{2}{*}{\(\quad\tau\simeq i\infty\quad\)} \\
 & & \(\,\,k+k^c \equiv 0 \,\,\)& \(\,\,k+k^c \equiv 1\,\, \) & \(\,\,k+k^c \equiv 2\,\, \) &\\
\midrule \endfirsthead
\caption{(cont.)}\\
  \toprule
\multirow{2}{*}{\(\quad\mathbf{r}\quad\)} &\multirow{2}{*}{\(\quad\mathbf{r}^c\quad\)} & & \(\quad\tau\simeq\omega\quad\) & & \multirow{2}{*}{\(\quad\tau\simeq i\infty\quad\)} \\
 & & \(\,\,k+k^c \equiv 0 \,\,\)& \(\,\,k+k^c \equiv 1\,\, \) & \(\,\,k+k^c \equiv 2\,\, \) &\\
\midrule \endhead
\bottomrule  \endfoot
%%%%%%%%%%%%
\(\mathbf{3}\) &
\(\mathbf{3}\) &
 \((1,1,1)\) &
 \((1,1,1)\) &
 \((1,1,1)\) &
 \((1,1,1)\) \\ 
%%%%%%%%%%%%
\(\mathbf{3}\) &
\(\mathbf{3'}\) &
 \((1,1,1)\) &
 \((1,1,1)\) &
 \((1,1,1)\) &
 \((1,1,\epsilon^2)\) \\ 
%%%%%%%%%%%%
\(\mathbf{3}\) &
\(\mathbf{\hat{3}}\) &
 \((1,1,1)\) &
 \((1,1,1)\) &
 \((1,1,1)\) &
 \((1,1,\epsilon^3)\) \\ 
%%%%%%%%%%%%
\(\mathbf{3}\) &
\(\mathbf{\hat{3}'}\) &
 \((1,1,1)\) &
 \((1,1,1)\) &
 \((1,1,1)\) &
 \((1,1,\epsilon)\) \\ 
%%%%%%%%%%%%
\(\mathbf{3'}\) &
\(\mathbf{3'}\) &
 \((1,1,1)\) &
 \((1,1,1)\) &
 \((1,1,1)\) &
 \((1,1,1)\) \\ 
%%%%%%%%%%%%
\(\mathbf{3'}\) &
\(\mathbf{\hat{3}'}\) &
 \((1,1,1)\) &
 \((1,1,1)\) &
 \((1,1,1)\) &
 \((1,1,\epsilon^3)\) \\ 
%%%%%%%%%%%%
\(\mathbf{\hat{3}}\) &
\(\mathbf{3'}\) &
 \((1,1,1)\) &
 \((1,1,1)\) &
 \((1,1,1)\) &
 \((1,1,\epsilon)\) \\ 
%%%%%%%%%%%%
\(\mathbf{\hat{3}}\) &
\(\mathbf{\hat{3}}\) &
 \((1,1,1)\) &
 \((1,1,1)\) &
 \((1,1,1)\) &
 \((1,1,\epsilon^2)\) \\
\end{longtable}

\renewcommand{\arraystretch}{1.2}
\begin{longtable}{cccccc}
\caption{
    Leading-order mass spectra patterns of bilinears \(\psi^c\psi\) in the vicinity of the symmetric points \(\omega\) and \(i \infty\), for 3d multiplets \(\psi \sim (\mathbf{r},k)\) and \(\psi^c \sim (\mathbf{r}^c,k^c)\) of the finite modular group \(\Gamma_5'\simeq A_5' = SL(2,\mathbb{Z}_5)\). Spectra are insensitive to transposition, i.e.~to the exchange \(\psi\leftrightarrow \psi^c\). Congruence relations for \(k+k^c\) are modulo 3.
}
\label{tab:A5p_patterns} \\
  \toprule
\multirow{2}{*}{\(\quad\mathbf{r}\quad\)} &\multirow{2}{*}{\(\quad\mathbf{r}^c\quad\)} & & \(\quad\tau\simeq\omega\quad\) & & \multirow{2}{*}{\(\quad\tau\simeq i\infty\quad\)} \\
 & & \(\,\,k+k^c \equiv 0 \,\,\)& \(\,\,k+k^c \equiv 1\,\, \) & \(\,\,k+k^c \equiv 2\,\, \) &\\
\midrule \endfirsthead
\caption{(cont.)}\\
  \toprule
\multirow{2}{*}{\(\quad\mathbf{r}\quad\)} &\multirow{2}{*}{\(\quad\mathbf{r}^c\quad\)} & & \(\quad\tau\simeq\omega\quad\) & & \multirow{2}{*}{\(\quad\tau\simeq i\infty\quad\)} \\
 & & \(\,\,k+k^c \equiv 0 \,\,\)& \(\,\,k+k^c \equiv 1\,\, \) & \(\,\,k+k^c \equiv 2\,\, \) &\\
\midrule \endhead
\bottomrule  \endfoot
%%%%%%%%%%%%
\(\mathbf{3}\) &
\(\mathbf{3}\) &
 \((1,1,1)\) &
 \((1,1,1)\) &
 \((1,1,1)\) &
 \((1,1,1)\) \\ 
%%%%%%%%%%%%
\(\mathbf{3}\) &
\(\mathbf{3'}\) &
 \((1,1,1)\) &
 \((1,1,1)\) &
 \((1,1,1)\) &
 \((1,\epsilon,\epsilon^4)\) \\ 
%%%%%%%%%%%%
\(\mathbf{3'}\) &
\(\mathbf{3'}\) &
 \((1,1,1)\) &
 \((1,1,1)\) &
 \((1,1,1)\) &
 \((1,1,1)\) \\ 
%%%%%%%%%%%%
\(\mathbf{3}\) &
\(\mathbf{1} \oplus \mathbf{1} \oplus \mathbf{1}\) &
 \((1,\epsilon,\epsilon^2)\) &
 \((1,\epsilon,\epsilon^2)\) &
 \((1,\epsilon,\epsilon^2)\) &
 \((1,\epsilon,\epsilon^4)\) \\ 
%%%%%%%%%%%%
\(\mathbf{3'}\) &
\(\mathbf{1} \oplus \mathbf{1} \oplus \mathbf{1}\) &
 \((1,\epsilon,\epsilon^2)\) &
 \((1,\epsilon,\epsilon^2)\) &
 \((1,\epsilon,\epsilon^2)\) &
 \((1,\epsilon^2,\epsilon^3)\) \\ 
%%%%%%%%%%%%
\(\mathbf{1} \oplus \mathbf{1} \oplus \mathbf{1}\) &
\(\mathbf{1} \oplus \mathbf{1} \oplus \mathbf{1}\) &
 \((1,1,1)\) &
 \((\epsilon^2,\epsilon^2,\epsilon^2)\) &
 \((\epsilon,\epsilon,\epsilon)\) &
 \((1,1,1)\) \\
\end{longtable}

\vfill
\clearpage

%%%%%%%%%%%%%%%%%%%%%%%
\bibliographystyle{JHEPwithnote}
\bibliography{bibliography}
%%%%%%%%%%%%%%%%%%%%%%%

\end{document}